\documentclass[11pt,letterpaper]{article}

\usepackage[utf8]{inputenc}
\usepackage[T1]{fontenc}
\usepackage{microtype}

\usepackage[margin=1in]{geometry}

\usepackage[skip=5pt plus 3pt minus 1pt,parfill=0pt]{parskip}
\makeatletter
\def\thm@space@setup{%
  \thm@preskip=\parskip \thm@postskip=0pt
}
\makeatother

\usepackage{amsmath,mathtools}
\usepackage{amsthm,thmtools}
\usepackage{amssymb}
\usepackage{bm,dsfont}
\usepackage{csquotes}

\usepackage{xcolor}

\usepackage{enumitem}

\usepackage[small]{caption}
\usepackage{subcaption}
\usepackage{tikz}
\usepackage{pgfplots}
\pgfplotsset{compat=newest}

\usepackage{booktabs,tabularew}
\usepackage[para]{threeparttable}

\usepackage[ruled,linesnumbered,noline]{algorithm2e}

\usepackage{array}
\usepackage{derivative}

\usepackage[
 backend=biber,
 style=alphabetic,
 citetracker,
 hyperref=auto,
 maxcitenames=10, 
 minalphanames=3,
 maxalphanames=4,
 sortcites,      
 sorting=nyt,
 maxbibnames=12, 
 date=year,      
 isbn=false,     
 url=false,      
 doi=false,      
 eprint=false,   
]{biblatex}
\newbibmacro{string+doiurl}[1]{%
  \iffieldundef{doi}{%
    \iffieldundef{url}{%
      #1
    }{%
      \href{\thefield{url}}{#1}%
    }%
  }{%
    \href{http://dx.doi.org/\thefield{doi}}{#1}%
  }%
}

\renewbibmacro*{volume+number+eid}{%
  \printfield{volume}%
  \setunit*{\adddot}%
  \printfield{number}}

\newbibmacro*{eid-or-pages}{%
  \iffieldundef{eid}
    {\printfield{pages}}%
    {\printfield{eid}}}

\renewbibmacro*{note+pages}{%
  \printfield{note}%
  \setunit{\bibpagespunct}%
  \usebibmacro{eid-or-pages}%
  \newunit}

\renewbibmacro*{chapter+pages}{%
  \printfield{chapter}%
  \setunit{\bibpagespunct}%
  \usebibmacro{eid-or-pages}%
  \newunit}

\AtEveryCitekey{%
  \ifciteseen
    {\defcounter{maxnames}{2}}
    {}%
}

\DeclareFieldFormat%
[article,inbook,incollection,inproceedings,patent,thesis,unpublished]
  {title}{\usebibmacro{string+doiurl}{#1}}

\DeclareFieldFormat{eid}{art.~#1}

\usepackage[colorlinks]{hyperref}
\hypersetup{citecolor=green!60!black,linkcolor=blue!70!black,urlcolor=blue!70!black}
\usepackage[capitalize, noabbrev, nameinlink]{cleveref}
\usepackage{xurl}
\hypersetup{breaklinks=true}

\usepackage{xspace}

\declaretheorem{theorem}
\declaretheorem{proposition}
\declaretheorem{lemma}

\declaretheorem{observation}

\declaretheorem{corollary}
\declaretheorem[style=definition]{definition}

\newcommand{\ind}{\ensuremath{\mathds{1}}}

\newcommand{\alg}{{\mathrm{ALG}}}
\newcommand{\opt}{{\mathrm{OPT}}}

\definecolor{bluegreen}{HTML}{33615E}
\definecolor{lime}{HTML}{56A778}
\definecolor{yellow}{HTML}{E5CB6D}
\definecolor{orange}{HTML}{FB8318}
\definecolor{lightred}{HTML}{DD6143}

\definecolor{job1}{HTML}{A6CEE3}
\definecolor{job2}{HTML}{1F78B4}
\definecolor{job5}{HTML}{B2DF8A}
\definecolor{job6}{HTML}{33A02C}
\definecolor{job3}{HTML}{FB9A99}
\definecolor{job4}{HTML}{E31A1C}

\usepackage[colorinlistoftodos,prependcaption,textsize=scriptsize]{todonotes}

\title{Tight Worst-Case Completion-Time Bounds for Dominant Resource Fairness}

\author{Mugen Blue\thanks{University of California, Santa Cruz, US. \texttt{msblue@ucsc.edu}} \and Sungjin Im\thanks{University of California, Santa Cruz, US. \texttt{sim9@ucsc.edu}} \and Alexander Lindermayr\thanks{Institut für Mathematik, Technische Universität Berlin, Germany. \texttt{alexander.lindermayr@tu-berlin.de}}}

\date{}

\bibliography{literature}

\begin{document}

\maketitle

\begin{abstract}

Dominant Resource Fairness (DRF) is an influential policy for sharing
multiple resources among jobs with heterogeneous demands. It combines
sharing incentive, strategy-proofness, envy-freeness, and Pareto
efficiency, making it particularly attractive for shared computing
systems. DRF has been implemented in major resource-management
frameworks, including Apache Mesos and Hadoop YARN. DRF is also
nonclairvoyant: it requires no advance knowledge of job sizes, making it
practical when processing requirements are difficult to predict. This
motivates understanding its completion-time performance relative to an
optimal schedule that knows all job sizes. However, elementary arguments
can only give an $\Omega(\sqrt D)$ lower bound and an $O(D)$
upper bound for total completion time, where $D$ is the number of
resource types.

In this paper, we close this gap by proving that DRF is
$\Theta(\sqrt D)$-competitive.
More generally, for every fixed real $q\geq1$, we show that DRF is
$\Theta(D^{1/(q+1)})$-competitive for the $\ell_q$ norm of completion
times, with larger $q$ placing greater emphasis on late-finishing jobs.
These bounds hold with arbitrary release times. We also give a refined
analysis of total completion time for small $D$.
 \end{abstract}

\section{Introduction}
\label{sec:introduction}

Shared computing systems must coordinate access to many resources,
including CPU, memory, accelerators, storage I/O, and network bandwidth.
In \emph{multidimensional scheduling}, each distinct resource type or
separately constrained resource pool corresponds to a distinct dimension,
and a job's demand vector specifies how much of each resource it needs
per unit of processing. The number of dimensions need not be limited to
a few hardware categories. For example, when jobs transmit data along
fixed network routes, each link imposes its own bandwidth constraint and
can be represented by a separate dimension. Thus, the dimension $D$ can
grow with the number of separately constrained resources in the system,
making it important to understand how scheduling performance scales with
$D$. Accounting for these demands together avoids the waste that fixed
resource bundles can cause when jobs have different bottlenecks, a
central motivation of Ghodsi et al.~\cite{ghodsi2011drf}.

In their influential paper, Ghodsi et al.~\cite{ghodsi2011drf} introduced
\emph{Dominant Resource Fairness} (DRF), which extends max-min fairness by comparing users through
their dominant shares: the largest fraction of any resource allocated to
each user. In their allocation model, DRF simultaneously satisfies four
properties particularly well suited to a shared system.
\emph{Sharing incentive} ensures that a user does at least as
well as with an equal private partition of the resources.
\emph{Strategy-proofness} rules out gains from misreporting resource
demands. \emph{Envy-freeness} means that no user prefers another user's
allocation. Finally, \emph{Pareto efficiency} means that improving one
user's allocation must hurt another. Together, these properties encourage
participation and truthful reporting, ensure fairness between users, and
rule out avoidable inefficiency. Among the allocation policies compared
in the original paper, DRF is the only one satisfying all four
properties~\cite[Table~2]{ghodsi2011drf}.

DRF has also influenced practical resource management: it has been
implemented in the Apache Mesos cluster resource
manager~\cite{mesos-drf} and is available as a built-in allocation policy
in Hadoop YARN's Fair Scheduler~\cite{hadoop-fair-scheduler}.
Its fractional allocation
has a direct progressive-filling implementation: increase dominant shares
together, freeze jobs when they encounter a binding constraint, and
continue with the remaining jobs~\cite{ghodsi2011drf,parkes2015beyond}.
We give the formal definition of DRF and its progressive-filling
algorithm in \cref{sec:model-and-algorithms}.
This simple structure makes DRF an
attractive policy to understand even when other allocation rules admit
stronger optimization guarantees. Our analysis concerns this fractional
scheduling model, rather than the placement and indivisibility constraints
of a particular cluster implementation.

Fairness of an instantaneous allocation does not, however, determine when
jobs finish. A policy can divide resources fairly at every moment and
still delay many jobs relative to a schedule that chooses a different
allocation over time. This distinction is especially important in the
\emph{nonclairvoyant} setting: the scheduler knows a job's resource demands
but learns its total processing requirement only when the job completes.
Here a natural benchmark is the total completion time of an optimal
offline schedule that knows all processing requirements in advance.

For \emph{Proportional Fairness} (PF), a classical allocation rule from
communication networks~\cite{kelly1998rate}, this connection between
allocation and scheduling is well understood. Im, Kulkarni, and
Munagala~\cite{im2018competitive} proved that PF is $O(1)$-competitive\footnote{An algorithm is said to be $c$-competitive if its objective is at most $c$ times the objective of the (offline) optimum solution for \emph{all} inputs.}
for total weighted completion time under general polyhedral packing
constraints, which include multidimensional scheduling. PF chooses rates
by maximizing a weighted sum of their logarithms. DRF instead has the
direct progressive-filling description above, without requiring a general
nonlinear optimization routine. The performance guarantee for PF does
not transfer to DRF: indeed, Im et al.~\cite{im2018competitive} already
observed that DRF need not be $O(1)$-competitive.

Elementary arguments give two bounds that leave a substantial gap.
With $D$ resource types, a simple instance gives an
$\Omega(\sqrt D)$ lower bound for total
completion time; we reproduce the construction in
\cref{prop:drf-lower-bound}. On the other hand, a straightforward comparison
gives an $O(D)$ upper bound; see \cref{thm:lq-dimension-upper-bound}.
Thus some dependence on the number of
resources is unavoidable for DRF, but its correct magnitude is not
determined by these bounds. This raises the question:
\begin{quote}
\emph{What is the asymptotically tight competitive ratio of DRF for job
completion times as a function of the number of resources?}
\end{quote}

\subsection{Our Results}

We resolve this question by proving that DRF is
$\Theta(\sqrt D)$-competitive for total completion time. More generally,
we give an asymptotically tight analysis for every $\ell_q$ norm of the
completion-time vector. The guarantees hold with arbitrary release times. The algorithm
knows demand vectors and caps but not job sizes or future arrivals.

For a completion-time vector $C=(C_j)_{j=1}^n$ and a real $q\geq1$, the
objective is
\[
    \lVert C\rVert_q=\left(\sum_{j=1}^n C_j^q\right)^{1/q}.
\]
When $q=1$, this is total completion time, equivalently average completion
time up to the factor $n$. Larger $q$ penalizes large completion times
more strongly, while the limit as $q\to\infty$ is the maximum completion
time. These objectives let us examine whether an allocation policy's
aggregate performance comes at the expense of particularly late-finishing
jobs.

\paragraph{Tight dependence on the dimension and norm.}
For every integer $D\geq1$ and real $q\geq1$, the competitive ratio of
DRF is in
\[
    \Theta\!\left(\min\left\{qD^{1/(q+1)},D\right\}\right).
\]
\Cref{thm:drf-norms,cor:drf-norms-release-times} establish this bound
for zero and arbitrary release times, respectively. The hidden constants
are independent of both $D$ and $q$.
In particular, for each fixed $q$, the dependence on the dimension is
$\Theta(D^{1/(q+1)})$.
The allocation rule itself does not depend on $q$. These are tight bounds
for DRF, not lower bounds for every nonclairvoyant algorithm: PF's
dimension-independent constant-factor guarantee for total completion time illustrates
the distinction.

\paragraph{Sharper bounds for total completion time.}
We analyze the case $q=1$ separately to obtain a simpler proof and better
constants. When all jobs are released at time zero, DRF is
$(4\sqrt D+1)$-competitive, while its competitive ratio is at least
$1+\sqrt D$~(\cref{thm:drf-completion-time,prop:drf-lower-bound}).
For two and three resources, a more detailed analysis improves
the upper bounds to $2+\sqrt3$ and $2+\sqrt7$, respectively
(\cref{thm:drf-two-resources,thm:drf-three-resources}). These
low-dimensional guarantees are relevant when jobs compete for only a
small number of resource types.

\paragraph{Arbitrary release times.}
We first analyze instances in which all jobs are available at time zero.
A black-box reduction then extends the bounds to arbitrary release times
at a cost of at most two~(\cref{thm:release-time-reduction}).
The reduction applies to any positive allocation
rule that depends only on the active set and does not decrease a remaining
job's rate when other jobs are removed. DRF itself is not monotone in this
sense, but its simplified variant one-shot DRF is, and DRF completes every
job no later than this variant~(\cref{lem:full-drf-dominates-one-shot}).
In particular, the total-completion-time guarantee becomes
$8\sqrt D+2$, with improvements to $4+2\sqrt3$ and $4+2\sqrt7$ for two
and three resources~(\cref{cor:drf-release-times}).
\subsection{Overview of the Analysis}

For intuition, scale each resource's supply to one and first consider
jobs with binary demand vectors in these units. Each job therefore uses
a nonempty subset of resources and needs one unit of each to run at unit
rate. Assume also that all jobs are released at time zero and that rate caps
do not restrict processing. We refer to
\cref{sec:model-and-algorithms} for the full definitions of DRF and
one-shot DRF.

In this special case, DRF initially increases every unfinished job's rate
equally until some resource becomes fully utilized. It freezes the rates
of jobs using that resource and continues increasing the rates of the
other jobs, repeating until no rate can be increased. \emph{One-shot DRF}
stops at the first such bottleneck. At time $t$, let $m(t)$ be the number
of jobs unfinished under one-shot DRF, and let $c(t)$ be the maximum number
of these jobs that demand any one resource. One-shot DRF gives every
unfinished job rate $1/c(t)$.
We analyze this simpler rule: a coupling shows that full DRF completes
every job no later, so the resulting upper bounds also hold for full
DRF~(\cref{lem:full-drf-dominates-one-shot}).

\paragraph{The elementary bounds.}
Since $c(t)\leq m(t)$, every unfinished job receives at least rate
$1/m(t)$. This is the rate these $m(t)$ jobs would each receive if they
shared a single unit-speed machine equally, ignoring their heterogeneous
resource requirements. Such equal sharing is Round Robin. Since one-shot
DRF also gives all unfinished jobs the same rate, and their total rate
$m(t)/c(t)$ is at least one, every job completes under one-shot DRF no later
than under this single-machine policy.
In contrast, every feasible schedule has aggregate processing rate at
most $D$, because each job uses at least one unit-capacity resource.
Round Robin's constant competitive ratio thus gives the elementary
$O(D)$ upper bound~(\cref{thm:lq-dimension-upper-bound}).
For the $\Omega(\sqrt D)$ lower bound, take one unit-size job per resource
and $\Theta(\sqrt D)$ unit-size jobs that each require every resource.
All resources become tight together, so DRF gives every job
rate $\Theta(1/\sqrt D)$ and incurs total completion time
$\Theta(D^{3/2})$. A better scheduler can first complete the
single-resource jobs in parallel and then process the remaining jobs
sequentially, for total completion time $O(D)$
(\cref{prop:drf-lower-bound}).

\paragraph{The key distinction: concentrated demand or rapid progress.}
The improvement comes from distinguishing between (i)~times when some
resource is demanded by a large fraction of unfinished jobs and (ii)~times
when no resource is so heavily shared. In case~(i), one-shot DRF is slowed
by a bottleneck that also constrains OPT. Importantly, all jobs still
unfinished at time $t$ have received the same amount of processing, so
each of their processing requirements is at least this amount. In
particular, each of the $c(t)$ jobs using the bottleneck requires at
least this much processing on that resource even in OPT.
In case~(ii), $c(t)$ is small relative
to $m(t)$, and one-shot DRF makes rapid aggregate progress at rate
$m(t)/c(t)$. OPT's aggregate rate, meanwhile, cannot exceed $D$.

\paragraph{Cauchy--Schwarz for total completion time.}
We make this distinction quantitative by weighting the instantaneous
completion-time cost $m(t)$ in two complementary ways. Motivated by
case~(i), we weight $m(t)$ by the fraction $c(t)/m(t)$ of unfinished jobs
that demand the bottleneck resource, obtaining $c(t)$.
As noted above, each bottleneck job requires at least the common amount of processing
already received by the unfinished jobs. Using these
lower bounds on processing requirements and the nestedness of the
unfinished sets, we bound $\int_0^\infty c(t)\,\mathrm{d}t$ by $O(1)\opt$
(\cref{lem:bottleneck-bound}).
Motivated by case~(ii), we instead weight $m(t)$ by one-shot DRF's
aggregate processing rate $m(t)/c(t)$, obtaining $m(t)^2/c(t)$.
Comparing this rate with OPT's aggregate capacity $D$, we bound
$\int_{\{t:c(t)>0\}}m(t)^2/c(t)\,\mathrm{d}t$ by $O(D) \cdot\opt$
(\cref{lem:effective-mass-bound}).
Since the two weights are reciprocals, Cauchy--Schwarz combines the bounds:
\begin{align*}
    \alg
    =\int_0^\infty m(t)\,\mathrm{d}t
    \leq\left(\int_{\{t:c(t)>0\}}\frac{m(t)^2}{c(t)}\,\mathrm{d}t\right)^{1/2}
       \left(\int_0^\infty c(t)\,\mathrm{d}t\right)^{1/2}
     =O(\sqrt D) \cdot \opt.
\end{align*}
For two and three resources, we obtain smaller constants by treating each
resource as a separate unit-speed machine. On each machine, a job's
processing requirement is the work it needs from that resource. We split
each job's completion cost among the machines using nonnegative weights
that sum to one for that job, avoiding double counting. The sum of the
optimal weighted completion costs on these machines then gives a lower
bound on $\opt$. Unlike the aggregate-capacity bound motivated by
case~(ii), which uses only the total capacity $D$, this bound accounts for
which jobs compete for each resource
(\cref{sec:low-dimensional-bounds}).

\paragraph{LP dual fitting for general norms.}
For the $\ell_q$ objective, the analogous identity is
$\sum_j C_j^q=\int_0^\infty qt^{q-1}m(t)\,\mathrm{d}t$.
We use a continuous-time linear programming relaxation of this objective
and construct dual solutions from the one-shot DRF schedule. By weak
duality, each feasible dual solution gives a lower bound on the optimal
sum of $q$th powers of completion times.
Fix a threshold $B\in[1,D]$ and call a time \emph{heavy} if
$c(t)\geq m(t)/B$, and \emph{light} otherwise. If heavy times contribute
more to the objective, we partition time into intervals, starting a new
interval whenever $c(t)$ drops below half its value at the start of the
current interval. For each interval, we choose a resource with the most
unfinished jobs just before the interval ends and use it to construct
the dual prices. If light times contribute
more, we assign the same price to all resources and exploit the high
aggregate processing rate. These two constructions yield competitive
bounds $O(qB^{1/q})$ and $O(qD/B)$, respectively. Choosing
$B=D^{q/(q+1)}$ balances them and gives $O(qD^{1/(q+1)})$.
The elementary processor-sharing comparison also gives an $O(D)$ bound
for every $q$. The matching lower bounds extend the construction for
total completion time described above to the $\ell_q$ objective
(\cref{thm:lq-lower-bounds}).

\paragraph{General demands, rate caps, and release times.}
For nonbinary demands, we replace the number of jobs using a resource by
the sum of their demands per unit of processing. For example, a job
whose demand on a resource is $0.2$ contributes $0.2$ to this sum rather
than one.

A job may also have a maximum useful resource allocation: once it
receives all the resources it requests, it runs at its maximum processing
rate, and giving it more resources does not make it faster. We represent
this maximum rate by $u_j$, the job's \emph{rate cap}. A job of size
$p_j$ therefore needs at least $p_j/u_j$ time even in OPT. We separate
this unavoidable delay from congestion. In the remaining analysis, a
cap-limited job counts fractionally, according to its actual rate
relative to the common rate. On each resource, we sum the jobs' demands
multiplied by these fractions. These weighted sums replace the binary
job counts in the arguments above; they are not the resource's current
utilization or the amount of work remaining on it.

Finally, consider arbitrary release times. Removing other active jobs
reduces competition for resources, so one-shot DRF can only increase a
remaining job's rate. Equivalently, if we include not-yet-released jobs
when computing the allocation at a given time, the rates assigned to
already active jobs cannot increase. We use this fact to
construct a virtual instance in which all jobs are released at time zero,
choosing their sizes so that the algorithm has the same completion times
as in the original instance. Each job's virtual size is at most its
original size plus the processing it receives before its original release
time. Combining a schedule for this extra early processing with an
optimal schedule for the original instance costs at most a factor of two
(\cref{thm:release-time-reduction}).

\subsection{Other Related Work}
\label{sec:related-work}

The product-maximization principle underlying PF dates back to
Nash~\cite{nash1950bargaining}. 
Kelly, Maulloo, and
Tan~\cite{kelly1998rate} developed PF for bandwidth sharing in communication networks.
The closest worst-case results concern PF in nonclairvoyant scheduling
under general packing constraints. As discussed above, Im, Kulkarni, and
Munagala~\cite{im2018competitive} proved a dimension-independent constant
competitive ratio for total weighted completion time in this model.
J\"ager, Lindermayr, and Megow~\cite{jaeger2026power} improved the bound
to $27$, and to $4$ when removing jobs cannot decrease the remaining
jobs' PF rates. This monotonicity holds, for example, in related-machine and
restricted-assignment scheduling. In these two models, they also obtain
a tight factor of $2$ for unweighted completion time with simultaneous
arrivals. These results do not transfer to DRF, whose
completion-time guarantees necessarily depend on the number of resources.

Subsequent work on DRF studies weighted entitlements and indivisible
resources~\cite{parkes2015beyond}, and incentives under changing demands
when allocations account for past service~\cite{fikioris2024incentives}.
Unlike DRF, PF is not strategy-proof in general. Cole, Gkatzelis, and
Goel~\cite{cole2013partial} construct a truthful mechanism that guarantees
each user at least a $1/e$ fraction of its PF utility for fixed-proportion
valuations. The mechanism withholds resources and thus gives up Pareto
efficiency. These works concern allocation and strategic behavior rather
than completion-time guarantees.

On a single unit-speed machine, processor sharing, or fractional
Round Robin (RR), is tightly $2$-competitive for total completion time
when all jobs arrive together~\cite{motwani1994nonclairvoyant,moseley2022roundrobin}.
For $\ell_q$ norms of completion times, Moseley and
Vardi~\cite{moseley2022roundrobin} prove a tight $(q+1)^{1/q}$ bound
with simultaneous arrivals and integer $q\geq1$, and a factor of $4$
with arbitrary release times and real $q\geq1$.
With arbitrary releases, RR's ratio for total completion time can exceed
$2$: J\"ager, Lindermayr, and Megow~\cite{jaeger2026power} give a lower
bound of $2.19$.
With a single resource and nonbinding rate caps, DRF reduces to RR.
Our work determines how these guarantees degrade with the number of
resources when jobs have heterogeneous resource demands.

A separate literature studies flow time, $C_j-r_j$, often with speed
augmentation, which gives the online algorithm faster resources than its
offline benchmark. See~\cite{kalyanasundaram2000speed,bansal2010server}
for single-machine results and~\cite{chekuri2004multiprocessor} for
identical parallel machines. Further results address unrelated
machines~\cite{anand2012dual,im2014selfishmigrate} and parallel jobs with
general speedup curves~\cite{edmonds2012scalably}.
The gradient-descent framework of Chen, Im,
and Petety~\cite{chen2025gradient} gives further results across several
scheduling models. These guarantees concern a different objective and
benchmark from our same-capacity completion-time bounds.
For broader background on online scheduling and resource augmentation,
see the survey by Pruhs, Sgall, and Torng~\cite{pruhs2004online}.

Fair sharing has also been studied extensively under stochastic
workloads. Bonald and Massouli\'e~\cite{bonald2001impact} analyze the
stability and throughput of fair bandwidth-sharing policies, while
Bonald et al.~\cite{bonald2006queueing} compare max-min fairness, PF, and
balanced fairness in processor-sharing networks.
Massouli\'e~\cite{massoulie2007structural} establishes PF stability under
more general service and routing models. Closer to our setting, Bonald
and Roberts~\cite{bonald2014multiresource} compare DRF and PF under
dynamic multiresource workloads, studying throughput and delay.
Such analyses address stability and steady-state performance under
probabilistic assumptions. Our competitive analysis instead compares
DRF with a clairvoyant optimum on every finite instance, without
distributional assumptions on arrivals or job sizes.

Our analysis also connects to LP-based dual fitting. Anand, Garg, and
Kumar~\cite{anand2012dual} apply this technique to weighted flow time with
resource augmentation. The PF analyses of Im, Kulkarni, and
Munagala~\cite{im2018competitive} and J\"ager, Lindermayr, and
Megow~\cite{jaeger2026power} use the optimality conditions of PF's
instantaneous allocation program to guide the construction of dual lower
bounds. For general $\ell_q$ norms, our dual variables instead exploit
DRF's bottleneck structure.

\subsection{Organization of the Paper}

\Cref{sec:model-and-algorithms} defines the model and the two DRF variants.
\Cref{sec:completion-time} establishes the tight asymptotic bound for total
completion time, and \cref{sec:low-dimensional-bounds} in the appendix
sharpens the upper bounds for two and three resources. \Cref{sec:lq-general-case}
gives the upper and lower bounds for general $\ell_q$ norms. Proofs omitted
from the main body are also collected in the appendix.
Finally, \cref{sec:release-time-reduction} in the appendix proves the
black-box reduction and states the resulting guarantees for arbitrary
release times.

\section{Model and Algorithms}
\label{sec:model-and-algorithms}

\subsection{Multidimensional Scheduling}

We consider $n$ jobs sharing $D$ divisible resources, such as CPU, memory,
or network bandwidth, each of unit capacity.
Job $j$ has a processing requirement $p_j>0$, a release time $r_j\geq0$, a rate cap
$u_j>0$, and a demand vector
\[
    a_j=(a_{j1},\ldots,a_{jD})\in[0,1]^D \; \text{ with } \; \lVert a_j\rVert_\infty=1.
\]
The entry $a_{jd}$ is the amount of resource $d$ that job $j$ uses per unit of
processing rate. The resource that job $j$ demands most is its
\emph{dominant resource}, and we normalize $a_j$ so that this entry equals one.
Unless stated otherwise, all jobs are released at time zero. We extend our
results to arbitrary release times in \cref{sec:release-time-reduction}.

A schedule specifies a \emph{rate vector}
$x(t)\in\mathbb{R}_{\geq0}^n$ at every time $t\geq0$. We set
$x_j(t)=0$ before job $j$ is released. Running job $j$ at rate $x_j(t)$
consumes $a_{jd}x_j(t)$ units of resource $d$. A schedule is feasible if it
satisfies the following three constraints.
\begin{itemize}
    \item \textbf{Rate caps.} $0\leq x_j(t)\leq u_j$ for every job $j$
    and time $t$.
    \item \textbf{Resource capacities.}
    $\sum_{j=1}^n a_{jd}x_j(t)\leq1$ for every resource $d$ and time $t$.
    \item \textbf{Job completion.} Every job $j$ has a finite completion time
    $C_j$, which is the time at which it has received its full processing
    requirement, that is,
    $\int_{r_j}^{C_j}x_j(s)\,\mathrm{d}s=p_j$.
\end{itemize}
We call a single vector $x\in\mathbb{R}_{\geq0}^n$ a \emph{feasible rate
vector} if it satisfies the rate caps and resource capacities.

Our objective is to minimize the total completion time
$\sum_{j=1}^n C_j$. More generally, for a real $q\geq1$, we study the
$\ell_q$ norm $\lVert C\rVert_q=(\sum_{j=1}^n C_j^q)^{1/q}$ of the
completion-time vector $C=(C_j)_{j=1}^n$, which is the total completion
time for $q=1$. For a given objective, we write $\opt$ for the value of an
optimal feasible schedule and $\alg$ for the value attained by the
algorithm under consideration.

At time $t$, let $A(t):=\{j\in[n]:r_j\leq t<C_j\}$ be the set of
\emph{active} jobs, which have been released but not yet completed. When all
release times are zero, this is simply the set $U(t):=\{j\in[n]:t<C_j\}$ of
unfinished jobs.

The online algorithms analyzed throughout this paper are
\emph{nonclairvoyant}. When job $j$ arrives, the algorithm learns its demand
vector $a_j$ and rate cap $u_j$, but not its processing requirement $p_j$.
The algorithm only learns that a job is done when it completes, and it has no
information about jobs that have not yet been released. This models cluster
schedulers, which typically know the resources a job requests but not how
long it will run. Each allocation decision is irrevocable.

We say that an algorithm is \emph{$\mu$-competitive} for an objective if
$\alg(I)\leq\mu\cdot\opt(I)$ for \emph{every} instance $I$. The
infimum over all such $\mu$ is the algorithm's \emph{competitive ratio} for
this objective.

\subsection{Dominant Resource Fairness and One-Shot DRF}

Since $\lVert a_j\rVert_\infty=1$, the rate $x_j$ of job $j$ equals the
fraction of its dominant resource that it receives, which is called its
\emph{dominant share}. Dominant Resource Fairness (DRF)~\cite{ghodsi2011drf}
aims to equalize the dominant shares of all active jobs. It computes an
allocation by progressive filling, as described in \cref{alg:full-drf}.
The rates of all jobs start at zero and increase at the same speed. A job is
frozen, meaning that its rate stays fixed from then on, once it reaches its
cap or once a resource it uses becomes tight. The remaining jobs keep
increasing their rates. Each iteration freezes at least one additional job,
so the procedure terminates with a feasible rate vector. Over time, DRF
recomputes this allocation whenever the active set changes, that is, it runs
every job $j\in A(t)$ at rate $x_j^{\mathrm{DRF}}(A(t))$. We refer to this
algorithm as \emph{full DRF} to distinguish it from the one-shot variant
introduced below.

\begin{algorithm}[H]
    \caption{Full DRF allocation}
    \label{alg:full-drf}
    \KwIn{Active set $A$, normalized demands $a_j$ with
    $\lVert a_j\rVert_\infty=1$, and caps $u_j$}
    \KwOut{Rate vector $x^{\mathrm{DRF}}(A)$}
    Set $x_j\gets0$ for every $j\in A$ and $F\gets A$\;
    \While{$F\neq\emptyset$}{
        Increase $x_j$ at the same speed for every $j\in F$ until a job
        reaches its cap or a resource becomes tight\;
        Remove every job $j$ with $x_j=u_j$ from $F$\;
        \ForEach{resource $d$ with $\sum_{j\in A}a_{jd}x_j=1$}{
            Remove every job $j$ with $a_{jd}>0$ from $F$\;
        }
    }
    \Return{$x$}\;
\end{algorithm}

We call the part of progressive filling until a tight resource first freezes
a job strictly below its cap the \emph{initial round}. If no such event
occurs, the initial round ends when all jobs reach their caps. A resource
that becomes tight when all jobs using it have reached their caps does not
end this round; the other jobs continue increasing their rates.
The initial round has a simple closed-form description. For any
nonempty set $A$ of active jobs, define its \emph{common fair-share rate}
$\lambda(A)$ as the largest $\lambda$ such that running every job $j\in A$ at
rate $\min\{u_j,\lambda\}$ respects all resource capacities. Formally,
\begin{equation}
    \lambda(A):=\sup\left\{\lambda\geq0:
        \sum_{j\in A}a_{jd}\min\{u_j,\lambda\}\leq1
        \text{ for every }d\in[D]\right\}.
    \label{eq:common-fair-share-rate}
\end{equation}
For the empty active set, we set $\lambda(\emptyset):=\infty$.
If all jobs in $A$ can run at their caps simultaneously without exceeding any
resource capacity, then $\lambda(A)=\infty$. At the end of the initial round,
job $j$ has rate $\min\{u_j,\lambda(A)\}$. Full DRF may increase the rates of
some jobs during subsequent rounds. Its final allocation therefore satisfies
$\min\{u_j,\lambda(A)\}\leq x_j^{\mathrm{DRF}}(A)\leq u_j$ for every
$j\in A$.

We also study \emph{one-shot DRF}, which stops progressive filling after the
initial round. It assigns every job $j\in A$ the rate
\begin{equation}
    x_j^{\mathrm{OS}}(A):=\min\{u_j,\lambda(A)\}.
    \label{eq:one-shot-drf-allocation}
\end{equation}

This allocation is feasible by the definition of $\lambda(A)$. Unlike full
DRF, it does not redistribute capacity among the remaining jobs after the
initial round. One-shot DRF may therefore leave some capacity
unused. However, it is easier to analyze, because every job's rate depends on
the active set only through the single number $\lambda(A)$.
When applying the rule over time, we abbreviate
$\lambda(t):=\lambda(A(t))$ and $x_j(t):=x_j^{\mathrm{OS}}(A(t))$ for active
jobs, and set $x_j(t)=0$ for all other jobs.

A key property of one-shot DRF is that the rate of a job can only decrease
when more jobs are active.

\begin{observation}[Active-set monotonicity of one-shot DRF]
\label{obs:one-shot-monotonicity}
If $A\subseteq B$, then
\[
    \lambda(A)\geq\lambda(B)
    \qquad\text{and}\qquad
    x_j^{\mathrm{OS}}(A)\geq x_j^{\mathrm{OS}}(B)
    \quad\text{for every }j\in A.
\]
\end{observation}

\Cref{obs:one-shot-monotonicity} holds because removing jobs only relaxes the
constraints in \eqref{eq:common-fair-share-rate}, so every $\lambda$ that is
feasible for $B$ is also feasible for $A$.

\begin{restatable}{lemma}{lemFullDrfDominatesOneShot}
\label{lem:full-drf-dominates-one-shot}
On every instance, every job completes under full DRF no later than under
one-shot DRF.
\end{restatable}

We prove \cref{lem:full-drf-dominates-one-shot} in \cref{app:model}.

\Cref{lem:full-drf-dominates-one-shot} allows us to analyze one-shot DRF
instead of full DRF, since every upper bound for one-shot DRF also holds for
full DRF. This simplifies our analysis considerably. All uncapped jobs of
one-shot DRF run at the same rate $\lambda(t)$, and when all jobs are released
at time zero, the active set only shrinks over time. By
\cref{obs:one-shot-monotonicity}, the rate of every job is then nondecreasing
over time. With arbitrary release times, new arrivals may decrease the rates
of existing jobs. In \cref{sec:release-time-reduction}, we show that the
monotonicity in \cref{obs:one-shot-monotonicity} still suffices to extend our
bounds to this setting, losing only a factor of two.

Full DRF does not have this monotonicity property when demands can be zero.
For example, take four jobs with unit rate caps that demand the resource sets
$\{1\}$, $\{1,2\}$, $\{2\}$, and $\{2\}$, respectively. Full DRF gives the
first job rate $2/3$, but once the fourth job completes, every job runs at
rate $1/2$. We therefore apply the release-time reduction to one-shot DRF,
and full DRF inherits the resulting bounds by
\cref{lem:full-drf-dominates-one-shot}.

The original DRF paper~\cite{ghodsi2011drf} uses unnormalized demand vectors
and unit rate caps. In \cref{app:normalization}, we show that this
formulation is equivalent to ours and that original DRF and
\cref{alg:full-drf} give every job the same resource allocation at every
time and the same completion time. We use the normalized formulation because
$x_j$ is then exactly job $j$'s dominant share, which makes both progressive
filling and our analysis easier to express.

\section{Total Completion Time}
\label{sec:completion-time}

We begin by optimizing the total completion time. This is the special case of $q=1$ of the more general $\ell_q$-norm objective studied afterwards in \cref{sec:lq-general-case}. This special case admits a simpler proof with a better performance guarantee.

\subsection{Upper Bound}

In the first part of \Cref{sec:completion-time}, we will prove the following theorem.

\begin{theorem}
\label{thm:drf-completion-time}
For every instance of multidimensional scheduling with $D\geq1$
resources and zero-release times, one-shot DRF and full DRF satisfy
\[
    \alg\leq(4\sqrt D+1) \cdot \opt\leq5\sqrt D \cdot \opt
\]
for the total completion-time objective. Therefore, both algorithms are
$O(\sqrt D)$-competitive for minimizing total completion time.
\end{theorem}

By \cref{lem:full-drf-dominates-one-shot}, it suffices to analyze one-shot
DRF rather than full DRF. Throughout this section, we fix an instance and the
schedule of one-shot DRF with completion times $(C_j)_{j=1}^n$.

The proof has three steps. First, we observe that every job $j$ needs at
least $p_j/u_j$ time in every schedule, since its rate never exceeds its cap.
The sum of these times is a lower bound on $\opt$. We show that one-shot DRF exceeds this bound only by a
quantity $Z_1$, which measures how long jobs are slowed down by congestion.
Second, we use the Cauchy--Schwarz inequality to split $Z_1$ into two
factors. Third, we bound each factor by comparing it with $\opt$. The first
comparison uses only the total capacity of all $D$ resources and loses a
factor of $O(D)$. The second uses the bottleneck resources of one-shot DRF and
loses only a constant factor. Together, this gives $Z_1=O(\sqrt D)\cdot\opt$.

\subsubsection{Notation and setup} \label{sec:notation-setup-analysis}

Recall that $U(t)=\{j:t<C_j\}$ is the set of unfinished jobs. Since all jobs
are released at time zero, the active set equals $U(t)$ and hence
$\lambda(t)=\lambda(U(t))$. Moreover, $\lambda(t)=\infty$ exactly when all
unfinished jobs run at their caps.

We use the following quantities.
\begin{itemize}[nosep]
    \item The \emph{congestion}\footnote{If all demand vectors are binary and no job is limited by its rate cap, then $c(t)$ is simply the maximum number of unfinished jobs that demand the same resource at time $t$.} is $c(t):=1/\lambda(t)$, where $c(t) := 0$ if $\lambda(t) = \infty$.
    \item The \emph{effective mass} of job $j$ is
    $z_j(t):=\ind[j\in U(t)] \cdot \min\{1,u_jc(t)\}$.
    \item The \emph{effective job count} is $m(t):=\sum_{j=1}^nz_j(t)$.
    \item The \emph{effective resource load} of resource $d$ is $c_d(t) := \sum_{j=1}^n a_{jd} z_j(t)$.
\end{itemize}
Intuitively, $z_j(t)$ is the rate of job $j$ relative to the common rate
$\lambda(t)$. It equals one if job $j$ runs at rate $\lambda(t)$, and it is
smaller than one if job $j$ is limited by its cap. Hence, $m(t)$ counts the
unfinished jobs, where jobs limited by their caps count only fractionally.

We now make this intuition precise.
For every unfinished job $j$ with $c(t)>0$, we have $\lambda(t)<\infty$ and
the rate rule in
\eqref{eq:one-shot-drf-allocation} gives
\begin{align}
    c(t)x_j(t)
    =c(t)\min\{u_j,\lambda(t)\}
    =\min\{u_jc(t),c(t)\lambda(t)\}
    =\min\{u_jc(t),1\}=z_j(t) \ .
    \label{eq:capped-q-effective-rate}
\end{align}
If $j\notin U(t)$, then
both $x_j(t)$ and $z_j(t)$ are zero. If $c(t)=0$, then $z_j(t)=0$ for every
job and $c(t)x_j(t)=0$. Thus
\eqref{eq:capped-q-effective-rate} holds for every job and time.

It follows by feasibility of $x$ that for every resource $d\in[D]$, we have
\[
    c_d(t)=\sum_{j=1}^n a_{jd}z_j(t)
    =c(t)\sum_{j=1}^n a_{jd}x_j(t)
    \leq c(t) \ .
\]
If $c(t)>0$, then $\lambda(t)$ is finite, so at least one
resource $d^*\in[D]$ is tight in \eqref{eq:common-fair-share-rate}. For this
resource, we have
\[
    c_{d^*}(t)
    =c(t)\sum_{j=1}^n a_{jd^*}x_j(t)
    =c(t).
\]
If $c(t)=0$, then every $z_j(t)$ and
therefore every $c_d(t)$ is zero. In summary, the congestion equals the
effective load of the most loaded resource, that is,
\begin{equation}
\label{eq:capped-q-effective-congestion}
    c(t)=\max_{d\in[D]}c_d(t).
\end{equation}

Before we continue, we argue that many of the considered quantities are monotone over time.

\begin{restatable}[Monotonicity]{lemma}{lemMonotonicity}
\label{clm:capped-q-monotonicity}
The functions $c(t)$, $z_j(t)$, $m(t)$, and $c_d(t)$ are nonincreasing and
constant between job completions. At a completion time $t$, their values
are computed using $U(t)$, which excludes every job that completes at time
$t$. All four functions are zero after a finite time.
\end{restatable}

\begin{proof}
The set of unfinished jobs $U(t)$ only shrinks, so $\lambda(t)$ is nondecreasing and $c(t)$ is
nonincreasing. Both factors in the definition of $z_j(t)$ are nonnegative
and nonincreasing. The same therefore holds for $z_j(t)$ and for the sums
defining $m(t)$ and $c_d(t)$. All these quantities depend only on $U(t)$.
Therefore, they remain unchanged while no job completes. When one or more
jobs complete at time $t$, those jobs do not belong to
$U(t)=\{j:t<C_j\}$, so the values at time $t$ are already the values for
the smaller unfinished set.

Finally, $1/n$ is a feasible candidate fair-share rate for every unfinished
set. Every unfinished job therefore receives rate at least
$\min\{u_j,1/n\}>0$ and finishes in finite time. Once every job has
completed, all four quantities are zero.
\end{proof}

\subsubsection{Accounting for rate caps}\label{sec:total-completion-time-caps}

We now separate the contribution of the rate caps from the rest of the
objective. Job $j$ cannot finish before time $p_j / u_j$ in any feasible
schedule. Thus,
\begin{align*}
    \mathrm{CAP}_1&:=\sum_{j=1}^n\frac{p_j}{u_j} \le \opt \ .
\end{align*}
The remaining part of the objective is captured by the integral of the
effective mass. Since $z_j(t)=1$ while job $j$ runs below its cap,
this integral is at least the time that job $j$ spends below its cap.
Using the definitions\footnote{The subscript $1$ indicates $q=1$ as we will generalize this notation later to general $\ell_q$ norms.}
\begin{align*}
    Z_{j,1}:=\int_0^{C_j}z_j(t)\,\mathrm{d}t \text{ for every job $j \in [n]$, }
   \quad \text{and} \quad 
    Z_1:=\sum_{j=1}^n Z_{j,1}
    =\int_0^\infty m(t)\,\mathrm{d}t \ ,
\end{align*}
we have the following separation lemma. 
\begin{restatable}{lemma}{lemEffectiveTimeDecomposition}
\label{lem:effective-time-decomposition}
For every job $j$, we have $C_j\leq p_j/u_j+Z_{j,1}$. Consequently,
\[
    \alg\leq\mathrm{CAP}_1+Z_1\leq\opt+Z_1.
\]
\end{restatable}

\begin{proof}
Because $\lambda(t)$ is nondecreasing, a job that reaches its cap stays at
that cap until it completes. Let $\tau_j$ be the first time job $j$ runs at
its cap, or $C_j$ if it completes without reaching its cap. Before
$\tau_j$, the job has $z_j(t)=1$, so
\[
    Z_{j,1}\geq\int_0^{\tau_j}z_j(t)\,\mathrm{d}t=\tau_j.
\]
After $\tau_j$, any remaining work takes at most $p_j/u_j$ time. Hence
\[
    C_j\leq\tau_j+\frac{p_j}{u_j}
        \leq Z_{j,1}+\frac{p_j}{u_j}.
\]
Summing over all jobs and using $\mathrm{CAP}_1\leq\opt$ proves the claim.
\end{proof}

It remains to bound $Z_1$. Since $m(t)=0$ whenever $c(t)=0$,
Cauchy--Schwarz gives
\begin{equation}
\label{eq:effective-cauchy-schwarz}
    Z_1^2
    =\left(\int_0^\infty m(t)\,\mathrm{d}t\right)^2
    \leq
    \left(\int_{\{t:\,c(t)>0\}}\frac{m(t)^2}{c(t)}\,\mathrm{d}t\right)
    \left(\int_0^\infty c(t)\,\mathrm{d}t\right).
\end{equation}
We bound the first factor using the total processing capacity of all
resources (\cref{lem:effective-mass-bound}), and the second using the
bottleneck resources of one-shot DRF (\cref{lem:bottleneck-bound}).

\subsubsection{Lower bound from aggregate processing capacity}
\label{subsec:effective-mass-bound}

The first lower bound ignores the detailed geometry of the demand vectors
and uses only their normalization.

\begin{restatable}{lemma}{lemEffectiveMassBound}
\label{lem:effective-mass-bound}
It holds that
\[
    \int_{\{t:\,c(t)>0\}}\frac{m(t)^2}{c(t)}\,\mathrm{d}t\leq2D \cdot \opt.
\]
\end{restatable}

\begin{proof}
Since $\lVert a_j\rVert_\infty=1$ and demands are nonnegative, every job
satisfies $\sum_{d=1}^D a_{jd}\geq1$. Thus every feasible rate vector $y(t)$
satisfies
\[
    \sum_{j=1}^n y_j(t)
    \leq \sum_{j=1}^n \sum_{d=1}^D a_{jd}y_j(t)
    =\sum_{d=1}^D\sum_{j=1}^n a_{jd}y_j(t)
    \leq D.
\]

We may therefore relax the instance to a single machine of speed $D$,
discard all rate caps, and retain the processing requirements $p_j$. The
optimal value of this relaxation is a lower bound on $\opt$.
Shortest-processing-time first is optimal for total completion time on this
relaxation. Thus,
\begin{equation}
\label{eq:single-machine-lower-bound}
    \opt
    \geq \frac{1}{D} \left(\sum_{j=1}^np_j
    +\sum_{j=1}^{n-1}\sum_{j'=j+1}^n\min\{p_j,p_{j'}\} \right).
\end{equation}

We next bound the integral in the statement of the lemma. Since
$m(t)=\sum_{j=1}^nz_j(t)$, expanding its square gives
\[
    m(t)^2=\sum_{j=1}^n z_j(t)^2
    +2\sum_{j=1}^{n-1}\sum_{j'=j+1}^nz_j(t)z_{j'}(t).
\]
We bound the individual-job terms and the pairwise terms separately.
For every job $j$, the identity $z_j(t)=c(t)x_j(t)$ in \eqref{eq:capped-q-effective-rate} and the bound
$0\leq z_j(t)\leq1$ imply, whenever $c(t)>0$, that
\[
    \frac{z_j(t)^2}{c(t)}=x_j(t)z_j(t)\leq x_j(t).
\]
Integrating and using that job $j$ receives exactly $p_j$ units of
processing over its execution, we obtain
\[
    \int_{\{t:\,c(t)>0\}}\frac{z_j(t)^2}{c(t)}\,\mathrm{d}t
    \leq\int_0^\infty x_j(t)\,\mathrm{d}t=p_j.
\]
For two distinct jobs $j$ and $j'$, the same reasoning gives
\[
    \frac{z_j(t)z_{j'}(t)}{c(t)}
    \leq x_j(t)  \text{ and }  \frac{z_j(t)z_{j'}(t)}{c(t)}
    \leq x_{j'}(t)
    \qquad\text{when }c(t)>0.
\]
Thus by taking the integral over time, we have
\[
    \int_{\{t:\,c(t)>0\}}\frac{z_j(t)z_{j'}(t)}{c(t)}\,\mathrm{d}t
    \leq\min\{p_j,p_{j'}\} .
\]
Substituting these bounds into the expansion of $m(t)^2$ and applying
\eqref{eq:single-machine-lower-bound} yields
\begin{align*}
    \int_{\{t:\,c(t)>0\}}\frac{m(t)^2}{c(t)}\,\mathrm{d}t
    \leq\sum_{j=1}^n p_j
      +2\sum_{j=1}^{n-1}\sum_{j'=j+1}^n\min\{p_j,p_{j'}\}
    \leq 2D \cdot \opt \ ,
\end{align*}
which completes the proof of the lemma.
\end{proof}

\subsubsection{Lower bound from bottleneck resource loads}
\label{subsec:bottleneck-bound}

We next bound $\int_0^\infty c(t)\,\mathrm{d}t$ in terms of $\opt$. We use the allocation rule of one-shot DRF to relate congestion to processing requirements on a bottleneck resource. These requirements limit how many jobs $\opt$ can complete by a given time.

For the proof of this bound, and for the refined bounds in
\cref{sec:low-dimensional-bounds}, we divide the schedule of one-shot DRF
into phases.
Partition the schedule at its job completion times and omit phases with
$c(t)=0$. Let $I_1,\ldots,I_r$ be the remaining phases in chronological
order, with $r=0$ if there are none. In phase $h$, write
$c_h:=c(t)$, $z_{j,h}:=z_j(t)$, and $m_h:=m(t)$, and define its
uncapped service amount $l_h:=|I_h|/c_h$, which is the processing a job
would receive during phase $h$ at the common rate without an individual
cap. By \eqref{eq:capped-q-effective-rate}, job $j$ actually receives
$l_hz_{j,h}$ units of processing during phase $h$. Therefore,
we have $\sum_{h=1}^r l_hz_{j,h}\leq p_j$
for every job $j$. The inequality may be strict because jobs can also be
processed during the omitted phases, in which they run at their caps.
Since \(m(t)=0\) in the omitted phases, we also have
$Z_1 = \int_0^\infty m(t)\,\mathrm{d}t = \sum_{h=1}^r l_hm_hc_h$.

\begin{restatable}{lemma}{lemBottleneckBound}
\label{lem:bottleneck-bound}
It holds that
\[
    \int_0^\infty c(t)\,\mathrm{d}t\leq8 \cdot \opt.
\]
\end{restatable}

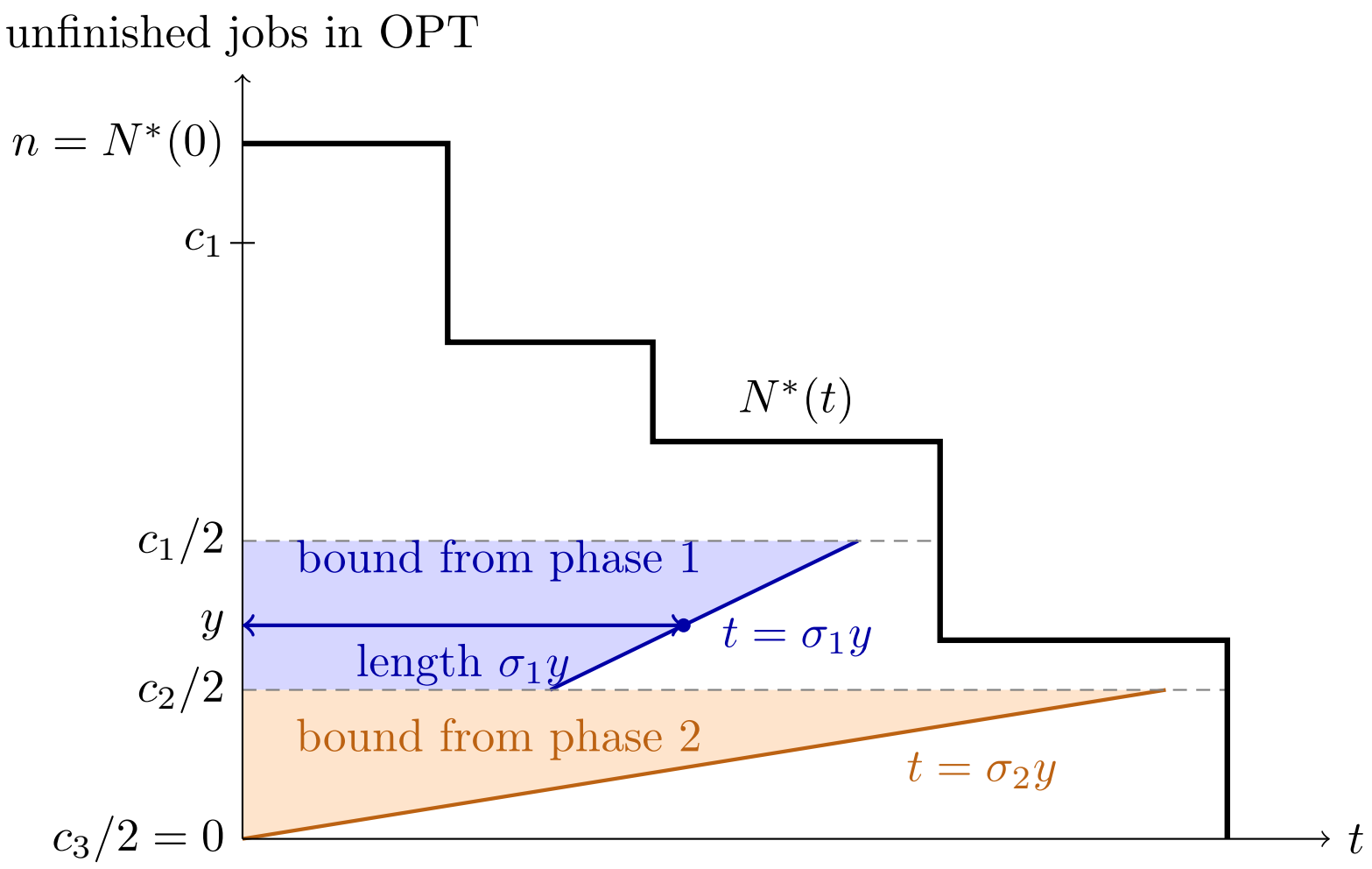
\begin{figure}[tbp]
\centering
\begin{tikzpicture}[x=1.55cm,y=.75cm, font=\small]
    \fill[blue!16] (0,1.5) -- (1.5,1.5) -- (3,3) -- (0,3) -- cycle;
    \fill[orange!22] (0,0) -- (4.5,1.5) -- (0,1.5) -- cycle;
    \draw[blue!65!black,thick] (1.5,1.5) -- (3,3);
    \draw[orange!75!black,thick] (0,0) -- (4.5,1.5);
    \draw[densely dashed,gray] (0,1.5) -- (4.8,1.5);
    \draw[densely dashed,gray] (0,3) -- (3.4,3);
    \draw[very thick] (0,7) -- (1,7) -- (1,5) -- (2,5)
        -- (2,4) -- (3.4,4) -- (3.4,2) -- (4.8,2) -- (4.8,0);
    \node[above] at (2.7,4) {$N^*(t)$};
    \draw[->] (0,0) -- (5.3,0) node[right] {$t$};
    \draw[->] (0,0) -- (0,7.7) node[above] {unfinished jobs in OPT};
    \node[left] at (0,0) {$c_3/2=0$};
    \node[left] at (0,1.5) {$c_2/2$};
    \node[left] at (0,3) {$c_1/2$};
    \node[left] at (0,6) {$c_1$};
    \draw (-0.06,6) -- (0.06,6);
    \node[left] at (0,7) {$n=N^*(0)$};
    \node[blue!65!black] at (1.25,2.8) {bound from phase $1$};
    \node[orange!75!black] at (1.25,1) {bound from phase $2$};
    \draw[<->,thick,blue!65!black] (0,2.15) -- (2.15,2.15)
        node[midway,below] {length $\sigma_1 y$};
    \node[left] at (0,2.15) {$y$};
    \fill[blue!65!black] (2.15,2.15) circle (1.5pt);
    \node[blue!65!black,below right] at (2.25,2.4) {$t=\sigma_1 y$};
    \node[orange!75!black,below right] at (3.15,1.05) {$t=\sigma_2 y$};
\end{tikzpicture}
\caption{Schematic of the area lower bound, with two phases and $c_3=0$.
The horizontal axis is time in OPT's schedule. The shaded regions are
bounds derived from DRF phases, not the time intervals of those phases.
Initially $N^*(0)=n\geq c_1$ (drawn with strict inequality). The unshaded area
above the bands is not counted in the lower bound.
At each height $y\in[c_{h+1}/2,c_h/2]$, phase $h$ provides a horizontal
segment $0\leq t\leq\sigma_hy$ below $N^*(t)$. Its shaded region has
area $\int_{c_{h+1}/2}^{c_h/2}\sigma_hy\,\mathrm{d}y$.
The two regions may overlap in time, but their height intervals are
disjoint, so their areas can be added without double counting.}
\label{fig:bottleneck-area}
\end{figure}

\begin{proof}
If $r=0$, then $c(t)=0$ at all times and the claim is immediate.
Otherwise, fix an optimal schedule $x^*$ and let $N^*(t)$ be the number of
jobs that are unfinished in this schedule at time $t$. Thus
$\opt=\int_0^\infty N^*(t)\,\mathrm{d}t$. Let
$\sigma_h:=\sum_{g=1}^h l_g$, with $\sigma_0:=0$, and set $c_{r+1}:=0$.
Since $c_h$ is nonincreasing by \Cref{clm:capped-q-monotonicity}, summation by parts gives
\begin{equation}
\label{eq:bottleneck-summation-by-parts}
    \int_0^\infty c(t)\,\mathrm{d}t
    =\sum_{h=1}^r l_hc_h^2
    =\sum_{h=1}^r\sigma_h\bigl(c_h^2-c_{h+1}^2\bigr).
\end{equation}

Fix a phase $h$. Some resource $d_h$ is tight in this phase, so
$\sum_{j=1}^na_{j d_h}z_{j,h}=c_h$. Moreover, the effective masses are
nonincreasing across phases by \Cref{clm:capped-q-monotonicity}, which implies
\[
    p_j\geq\sum_{g=1}^h l_gz_{j,g}
        \geq\sigma_hz_{j,h}
\]
for every job $j$.
We first use this phase to bound how many jobs remain unfinished in OPT.
To complete job $j$, OPT must perform $p_j\geq\sigma_hz_{j,h}$ units
of processing. Each unit of processing consumes $a_{j d_h}$ units of
resource $d_h$, so completing this job requires at least
$\sigma_ha_{j d_h}z_{j,h}$ units of that resource.

Fix any $y\in[0,c_h/2]$, and let $F$ be the jobs that OPT has completed
by time $\sigma_hy$. Since resource $d_h$ has capacity one, OPT can
have used at most $\sigma_hy$ units of this resource by that time. Hence
\[
    \sigma_h\sum_{j\in F}a_{j d_h}z_{j,h}
    \leq\sum_{j\in F}p_ja_{j d_h}
    \leq\sigma_hy,
    \qquad\text{so}\qquad \sum_{j\in F}a_{j d_h}z_{j,h}\leq y.
\]
Since $\sum_{j=1}^na_{j d_h}z_{j,h}=c_h$, the jobs unfinished in OPT satisfy
\[
    \sum_{j\in[n]\setminus F}a_{j d_h}z_{j,h}\geq c_h-y.
\]
Note that the quantities $a_{j d_h}z_{j,h}$ are determined by phase $h$ of
one-shot DRF and not by the remaining work in OPT. Each of them is at most
one, so OPT must have at least $c_h-y$ unfinished jobs at time $\sigma_hy$. Because $y\leq c_h/2$, we conclude that
\[
    N^*(\sigma_hy)\geq c_h-y\geq y.
\]
All jobs are released at time zero, so $N^*(t)$ is nonincreasing.
Thus $N^*(t)\geq y$ throughout $0\leq t\leq\sigma_hy$.

We now combine these bounds across phases without double counting.
Recall that $\opt$ is the area under the curve $N^*(t)$.
The bound just proved says that, at height $y\leq c_h/2$, this area
contains a horizontal segment of length $\sigma_hy$.
For phase $h$, we count only the segments at heights
$y\in[c_{h+1}/2,c_h/2]$. Since $c_h$ is nonincreasing, these height
intervals have disjoint interiors for different phases. No area is
therefore counted twice. See \cref{fig:bottleneck-area} for a visualization.
Integrating the segment lengths over their
assigned heights and using
\eqref{eq:bottleneck-summation-by-parts} gives
\begin{align*}
    \opt
    \geq\sum_{h=1}^r
        \int_{c_{h+1}/2}^{c_h/2}\sigma_hy\,\mathrm{d}y
    =\frac18\sum_{h=1}^r\sigma_h\bigl(c_h^2-c_{h+1}^2\bigr)
     =\frac18\int_0^\infty c(t)\,\mathrm{d}t.
\end{align*}
This completes the proof of the lemma.
\end{proof}

\subsubsection{Completing the upper bound}

\begin{proof}[Proof of \cref{thm:drf-completion-time}]
Combining \eqref{eq:effective-cauchy-schwarz} with
\cref{lem:effective-mass-bound,lem:bottleneck-bound} gives
\[
    Z_1\leq\sqrt{2D \cdot \opt\cdot8\opt}=4\sqrt D\,\opt.
\]
\Cref{lem:effective-time-decomposition} therefore yields
\[
    \alg\leq\mathrm{CAP}_1+Z_1
        \leq(4\sqrt D+1)\cdot \opt
        \leq5\sqrt D\cdot \opt.
\]
This proves the bound for one-shot DRF when all jobs are released at time
zero. A common positive release time adds the same amount to every job's
completion time in both schedules, so the bound continues to hold.
Finally, by \cref{lem:full-drf-dominates-one-shot}, full DRF completes every
job no later than one-shot DRF and satisfies the same bound.
\end{proof}

In \cref{sec:release-time-reduction}, we remove the assumption that all jobs
are released at time zero, losing only a factor of two.

\subsection{Lower Bound}

We now show that the dependence on $D$ is tight up to a constant factor.
The instance combines jobs that each use a single resource with jobs that
use every resource. Both algorithms give all jobs the same rate, whereas a
comparison schedule first completes the single-resource jobs in parallel.

\begin{proposition}
\label{prop:drf-lower-bound}
For every integer $D\geq1$, the competitive ratio of one-shot DRF and full
DRF for minimizing total completion time in multidimensional scheduling
with $D$ resources is at least $1+\sqrt D$, even for
binary demands, release times $r_j=0$, and $p_j=u_j=1$.
\end{proposition}

\begin{proof}
Fix an integer $m\geq1$. For each resource $d\in[D]$, create $m$ jobs that
demand only resource $d$. Also create $k:=\lfloor m\sqrt D\rfloor$ jobs
that demand all resources. Every positive demand is one, and every job has
$p_j=u_j=1$ and $r_j=0$.

Each resource is demanded by exactly $m+k$ jobs, so one-shot DRF assigns
every job rate $1/(m+k)$. All resources are tight at this rate, so full DRF
makes the same allocation. All jobs complete at time $m+k$, giving
\[
    \alg=(Dm+k)(m+k).
\]
We compare this with a schedule that first processes the single-resource
jobs in $m$ rounds, one job per resource in parallel, and then processes
the remaining $k$ jobs sequentially. Thus
\[
    \opt\leq\frac{Dm(m+1)}2+mk+\frac{k(k+1)}2.
\]
The first term is the contribution of the single-resource jobs. The
remaining jobs complete at times $m+1,\ldots,m+k$, giving the last two
terms. Since $k/m\to\sqrt D$ as $m\to\infty$, we obtain
\[
    \frac{\alg}{\opt}
    \geq\frac{2(Dm+k)(m+k)}{Dm(m+1)+2mk+k(k+1)}
    \longrightarrow 1+\sqrt D.
\]
\end{proof}

Together with \cref{thm:drf-completion-time}, this shows that both algorithms
have competitive ratio $\Theta(\sqrt D)$ for total completion time with
zero-release times.

\section{Minimizing \texorpdfstring{$\ell_q$}{ell-q} Norms}
\label{sec:lq-general-case}

In this section, we study the $\ell_q$-norm objective when all jobs are
released at time zero.
Given a real number $q \ge 1$ and a feasible schedule with completion times $C=(C_j)_{j=1}^n$,
the $\ell_q$ norm is $\| C \|_q = \bigl(\sum_{j=1}^n C_j^q\bigr)^{1/q}$.
For $q=1$, this is the total completion time. Larger values of $q$ put more
weight on jobs with large completion times, and the $\ell_q$ norm approaches
the maximum completion time as $q\to\infty$. Hence, $\ell_q$ norms
interpolate between average and worst-case job latency.
Our goal is to prove the following theorem.

\begin{theorem}\label{thm:drf-norms}
	For every integer $D\geq1$ and real $q\geq1$, one-shot DRF and full DRF
	have a competitive ratio in
	\[
	 \Theta\bigl(\min\bigl\{qD^{1/(q+1)},D\bigr\}\bigr)
	\]
	for minimizing the $\ell_q$ norm of completion times in multidimensional
	scheduling with $D$ resources and zero-release times.
	In particular, the competitive ratio is in $\Theta(D)$ if $q \ge D$.
\end{theorem}

For $q=1$, the theorem recovers the $\Theta(\sqrt D)$ bound of
\cref{sec:completion-time}.
We first prove the upper bound $O(qD^{1/(q+1)})$ in \cref{sec:upper-bound-q}
and then the upper bound $O(D)$ in \cref{sec:lq-dimension-upper-bound}.
Matching lower-bound instances for one-shot DRF and full DRF are given in
\cref{sec:lq-lower-bounds}, and the remaining omitted proofs of this section
in \cref{app:lq-general-case}.
The black-box extension to arbitrary release times is deferred to
\cref{sec:release-time-reduction}.

Throughout this section, we reuse the congestion $c(t)$, the effective
masses $z_j(t)$, the effective job count $m(t)$, and the effective resource
loads $c_d(t)$ from \cref{sec:notation-setup-analysis}.

\subsection{An Upper Bound Depending on \texorpdfstring{$q$}{q}}\label{sec:upper-bound-q}

Fix a real number $q \ge 1$ and assume that all jobs are released at time zero.
Now fix a schedule of one-shot DRF with completion times $(C_j)_{j=1}^n$.
Let $\alg_q := \| C \|_q$.

For any $s\in(0,1]$, a \emph{speed-$s$ schedule} has capacity $s$ on every
shared resource and rate cap $su_j$ for each job $j$.
Thus its rates $y_j(t)$ must satisfy, at every time $t\geq0$,
\begin{alignat*}{2}
    0\leq y_j(t)&\leq su_j &\quad&\text{for every }j\in[n],\\
    \sum_{j=1}^n a_{jd}y_j(t)&\leq s &&\text{for every }d\in[D].
\end{alignat*}
Let $\opt_{s,q}$ denote the minimum $\ell_q$ norm among speed-$s$ schedules.
In particular, $\opt_{1,q}$ is the usual unit-speed optimum. Speed-$s$
schedules only serve as an analysis tool. In the proof of
\cref{thm:lq-q-upper-bound}, we show that $\opt_{s,q}\leq\opt_{1,q}/s$, so
our final bounds compare against the usual unit-speed optimum.

The proof first separates the delay caused by rate caps. We then introduce
a continuous-time relaxation and construct two dual solutions to bound
the remaining contribution, depending on
whether the effective job count is concentrated on one resource or spread
across resources. Intuitively, the effective count discounts jobs whose rates
are already limited by their caps, so its weighted integral captures the part
of the objective caused by competition for shared resources. At heavy times,
the most loaded resource carries a significant fraction of this count,
allowing us to charge the delay to a single bottleneck. At light times, the count is spread
across resources, and we instead charge it through all $D$ resource
constraints. A threshold $B$ balances the losses in these two arguments and
leads to the factor $D^{1/(q+1)}$.

At the end, we use \cref{lem:full-drf-dominates-one-shot} to transfer the
guarantee to full DRF.

\subsubsection{Accounting for rate caps}

For every time $t\geq0$, we define the nondecreasing weight $\omega(t):=qt^{q-1}$. When $q=1$, we set $\omega(t)=1$ also at $t=0$.
For every time horizon $T\geq0$, we have
\begin{equation}
\label{eq:capped-q-weight}
    \int_0^T\omega(t)\,\mathrm{d}t
    =\int_0^T qt^{q-1}\,\mathrm{d}t
    =\bigl[t^q\bigr]_0^T
    =T^q.
\end{equation}
This identity lets us express completion-time powers as integrals and
separate the cap contribution from the rest of the objective.

We now separate the contribution of the rate caps, as in \Cref{sec:total-completion-time-caps}. Job $j$
cannot finish before time $p_j / (su_j)$
in a speed-$s$ schedule. Since $s \in (0,1]$, we have $p_j / (su_j) \ge p_j / u_j$ and thus
\begin{equation}
\label{eq:lb-q-opt-ineq}
    \mathrm{CAP}_q:=\biggl(\sum_{j=1}^n \biggl(\frac{p_j}{u_j}\biggr)^q\biggr)^{1/q}
    \leq\opt_{s,q}.
\end{equation}

As before, we capture the remaining part of the objective by integrating the
effective masses, now weighted by $\omega(t)$. For every job $j \in [n]$,
define
\[
    Z_{j,q}:=\int_0^{C_j}\omega(t)z_j(t)\,\mathrm{d}t
\]
and
\[
    Z_q:=\sum_{j=1}^n Z_{j,q}
    =\sum_{j=1}^n \int_0^{C_j}\omega(t)z_j(t)\,\mathrm{d}t
        = \int_0^\infty\omega(t) \sum_{j=1}^n z_j(t)\,\mathrm{d}t
        =\int_0^\infty\omega(t)m(t)\,\mathrm{d}t.
\]
For $q=1$, these are exactly the quantities $Z_{j,1}$ and $Z_1$ from
\cref{sec:total-completion-time-caps}.
The following separation lemma generalizes \Cref{lem:effective-time-decomposition}.

\begin{restatable}{lemma}{lemCappedQEffectiveDecomposition}
\label{lem:capped-q-effective-decomposition}
For every job $j$, let $\tau_j$ be the first time it runs at its cap, or
$C_j$ if it finishes without reaching its cap, and write
$\tau:=(\tau_j)_{j=1}^n$. Then
\[
    C_j\leq\frac{p_j}{u_j}+\tau_j
        \leq\frac{p_j}{u_j}+Z_{j,q}^{1/q}.
\]
Consequently, for every $s\in(0,1]$, we have
\[
    \alg_q\leq\opt_{s,q}+\lVert\tau\rVert_q
        \leq\opt_{s,q}+Z_q^{1/q}.
\]
\end{restatable}

\subsubsection{A continuous-time relaxation}

By \cref{lem:capped-q-effective-decomposition}, it remains to bound
$Z_q^{1/q}$ in terms of $\opt_{s,q}$. We do this by dual fitting. We
formulate a linear programming relaxation for speed-$s$ schedules and
construct feasible solutions to its dual from the schedule of one-shot DRF.
By weak duality, the objective value of every such dual solution is a lower
bound on $\opt_{s,q}^q$. This way, we never need to reason about the
structure of an optimal schedule directly.

Fix $s\in(0,1]$. In the following relaxation, the variable $y_j(t)$ is the
rate of job $j$ at time $t$.
\begin{alignat*}{2}
    \text{(LP)}\quad\min\quad
    &\sum_{j=1}^n\frac1{p_j}\int_0^\infty t^qy_j(t)\,\mathrm{d}t\\
    \text{s.t.}\quad
    &\frac1{p_j}\int_0^\infty y_j(t)\,\mathrm{d}t\geq1
        &\quad&\text{for every }j\in[n],\\
    &\sum_{j=1}^n a_{jd}y_j(t)\leq s
        &&\text{for every }d\in[D],\ t\geq0,\\
    &0\leq y_j(t)\leq su_j
        &&\text{for every }j\in[n],\ t\geq0.
\end{alignat*}
The objective charges each unit of processing of job $j$ at time $t$ a cost
of $t^q/p_j$. Since job $j$ receives $p_j$ units of processing before $C_j$,
its total cost in a feasible schedule is at most $C_j^q$.

Its continuous dual is given below, with dual variables $\alpha_j$,
$\beta_d(t)$, and $\gamma_j(t)$.
\begin{align*}
    \text{(DLP)}\quad\max\quad
    &\sum_{j=1}^n\alpha_j
      -s\int_0^\infty\sum_{d=1}^D\beta_d(t)\,\mathrm{d}t
      -s\sum_{j=1}^n u_j\int_0^\infty\gamma_j(t)\,\mathrm{d}t\\
    \text{s.t.}\quad
    &\frac{\alpha_j-t^q}{p_j}
      \leq\sum_{d=1}^D a_{jd}\beta_d(t)+\gamma_j(t)
        &&\text{for every }j\in[n],\ t\geq0,\\
    &\alpha_j\geq0 &&\text{for every }j\in[n],\\
    &\beta_d(t)\geq0 &&\text{for every }d\in[D],\ t\geq0,\\
    &\gamma_j(t)\geq0 &&\text{for every }j\in[n],\ t\geq0.
\end{align*}
We will always set $\gamma_j(t)=0$. The dual then has a natural economic
interpretation. We can think of $\beta_d(t)$ as the price of resource $d$ at
time $t$ and of $\alpha_j$ as a budget of job $j$. The dual constraint says
that the budget of job $j$ is at most the cost $t^q$ of completing at time
$t$ plus the price $p_j\sum_{d=1}^Da_{jd}\beta_d(t)$ of processing the
entire job at time $t$. The dual objective is the total budget minus the
total value of the resource capacity. We will choose the prices according to
the congestion in the schedule of one-shot DRF.

We first show that (LP) is indeed a relaxation of $\opt_{s,q}^q$.

\begin{restatable}{lemma}{lemCappedQPrimalRelaxation}
\label{lem:capped-q-primal-relaxation}
For every $q\geq1$ and $s\in(0,1]$, the optimal value of (LP) is at most
$\opt_{s,q}^q$.
\end{restatable}

The next lemma states weak duality with the cap variables $\gamma_j(t)$ set to zero.

\begin{restatable}[Weak-duality bound]{lemma}{lemCappedQDualBound}
\label{lem:capped-q-dual-bound}
Fix $s\in(0,1]$. Suppose that $\alpha_j\geq0$ for every $j\in[n]$ and that
each $\beta_d$ is a nonnegative and integrable function on
$[0,\infty)$. If
\begin{equation}
\label{eq:capped-q-dual-certificate}
    \alpha_j\leq t_0^q+p_j\sum_{d=1}^D a_{jd}\beta_d(t_0)
\end{equation}
for every $j\in[n]$ and $t_0\geq0$, then
\[
    \sum_{j=1}^n\alpha_j-s\int_0^\infty\sum_{d=1}^D\beta_d(t)\,\mathrm{d}t
    \leq\opt_{s,q}^q.
\]
\end{restatable}

\subsubsection{Dual fitting}

We now use \cref{lem:capped-q-dual-bound} to bound $Z_q$.
If $Z_q=0$, \cref{lem:capped-q-effective-decomposition} gives
$\alg_q\leq\opt_{1,q}$, so assume $Z_q>0$. Fix a parameter $B\in[1,D]$ which we will optimize later.

\begin{definition}[Heavy and light times]
A time $t\geq0$ is \emph{heavy} if $c(t)\geq m(t)/B$, and \emph{light}
otherwise.
\end{definition}

Denote the heavy-time and light-time contributions to $Z_q$ by
\begin{align*}
    Z_q^{\mathrm H}&:=\int_{\{t:\,t\text{ is heavy}\}}
        \omega(t)m(t)\,\mathrm{d}t \quad \text{and}\\
    Z_q^{\mathrm L}&:=\int_{\{t:\,t\text{ is light}\}}
        \omega(t)m(t)\,\mathrm{d}t.
\end{align*}
Then $Z_q=Z_q^{\mathrm H}+Z_q^{\mathrm L}$.
We distinguish two cases, depending on whether heavy or light times
contribute more to $Z_q$, and construct a different dual solution for each
case.

\subsubsection{Dual fitting case 1: heavy times dominate}

Assume that $Z_q^{\mathrm H}\geq Z_q^{\mathrm L}$. 
Let $T:=\min\{t\geq0:c(t)=0\}$
be the first time at which all unfinished jobs can run at their rate caps
simultaneously. 
Since $Z_q>0$, we have $c(0)>0$, so $T>0$. By
\cref{clm:capped-q-monotonicity}, $c(t)$ is
positive for $0\leq t<T$ and zero for all $t\geq T$.
Some jobs may continue running at their caps after $T$.

We next partition $[0,T)$ into intervals $I_i=[T_{i-1},T_i)$ as follows. Start at
$T_0=0$. For each interval, set $c_i:=c(T_{i-1})$ and let $T_i$ be the
first time $c(t)<c_i/2$. Such a time exists no later than $T$. Stop when
$T_i=T$, and let $k$ be the number of intervals. Thus
\begin{alignat*}{3}
    \frac{c_i}{2}\leq c(t)&\leq c_i
        &\quad&\text{for every }i\in[k],\ t\in I_i, \text{ and }\\
    c_{i+1}&\leq\frac{c_i}{2}
        &&\text{for every }i\in\{1,\ldots,k-1\}.
\end{alignat*}

For each interval $I_i$, consider the effective resource loads immediately
before its endpoint $T_i$. Taking the values immediately before $T_i$ in
\eqref{eq:capped-q-effective-congestion} gives
$c(T_i^-)=\max_{d\in[D]}c_d(T_i^-)$.
Choose a resource $d_i\in[D]$ attaining this maximum, and call it the
\emph{bottleneck resource} of $I_i$. We use the values
immediately before $T_i$ because jobs completing at $T_i$ are removed from
$U(T_i)$ and may cause the effective loads to decrease at that time.

Fix a time $t\in I_i$. Since $c(t)\leq c_i$, every unfinished job
$j\in U(t)$ satisfies
\[
    z_j(t)=\min\{1,u_jc(t)\}\leq\min\{1,u_jc_i\}.
\]
Moreover, \cref{clm:capped-q-monotonicity} implies that the effective load
on resource $d_i$ cannot increase between $t$ and $T_i$. Finally,
$c(T_i^-)\geq c_i/2$ by the definition of $I_i$. Combining these facts gives
\begin{equation}
\label{eq:capped-q-selected-load}
    \sum_{j\in U(t)}a_{j,d_i} \cdot \min\{1,u_jc_i\}
    \geq c_{d_i}(t)
    \geq c_{d_i}(T_i^-)
    =c(T_i^-)
    \geq\frac{c_i}{2}.
\end{equation}

The next lemma performs the full dual fitting argument and concludes this case.

\begin{restatable}[Heavy-case dual fitting]{lemma}{lemCappedQHeavyDualFitting}
\label{lem:capped-q-heavy-dual-fitting}
If $Z_q^{\mathrm H}\geq Z_q^{\mathrm L}$,
then
$\alg_q\leq\bigl(1+(8B)^{1/q}\bigr)\opt_{1/(8q),q}$.
\end{restatable}

\subsubsection{Dual fitting case 2: light times dominate}

Now assume that $Z_q^{\mathrm L}\geq Z_q^{\mathrm H}$. For every time $t\geq0$, define
$M(t):=\sup_{\theta\geq t}\omega(\theta)m(\theta)$.
Thus $M(t)$ bounds the weighted effective job count at every time at or after $t$.
It is nonincreasing because the set of later times only shrinks. It is
finite, and it is zero once $m$ has become zero.
In contrast to the heavy case, we now put the same price $M(t)/B$ on all
resources.

\begin{restatable}[Light-case dual fitting]{lemma}{lemCappedQLightDualFitting}
\label{lem:capped-q-light-dual-fitting}
If $Z_q^{\mathrm L}\geq Z_q^{\mathrm H}$,
then $\alg_q\leq\bigl(1+4^{1/q}\bigr)\opt_{B/(4qD),q}$.
\end{restatable}

\subsubsection{Proof of an upper bound depending on \texorpdfstring{$q$}{q}}

We now conclude \Cref{sec:upper-bound-q} by showing the following theorem.

\begin{theorem}
\label{thm:lq-q-upper-bound}
	For every integer $D\geq1$ and real $q\geq1$, one-shot DRF and full DRF
	have a competitive ratio in $O(q D^{1/(q+1)})$
	for minimizing the $\ell_q$ norm of completion times in multidimensional
	scheduling with $D$ resources and zero-release times.
\end{theorem}

\begin{proof}
Assume that all jobs are released at time zero.
Fix $s\in(0,1]$. Take an optimal unit-speed schedule with rates $y_j^*(t)$
and replace them by $s y_j^*(st)$. At time $t$, every resource load and job
rate is $s$ times its value at time $st$ in the original schedule. The
new schedule therefore respects resource capacities $s$ and rate caps
$su_j$. Moreover, for every job $j$ and time $t\geq0$,
\[
    \int_0^{t/s}s y_j^*(s\theta)\,\mathrm{d}\theta
    =\int_0^t y_j^*(\theta)\,\mathrm{d}\theta.
\]
Thus each completion time is multiplied by $1/s$, and so is the
$\ell_q$ norm. Since this is a feasible speed-$s$ schedule,
\[
    \opt_{s,q}\leq\frac1s\opt_{1,q}.
\]

If $Z_q=0$, \cref{lem:capped-q-effective-decomposition} with $s=1$ already
gives $\alg_q\leq\opt_{1,q}$. Hence, assume $Z_q>0$ and fix $B\in[1,D]$.
When $Z_q^{\mathrm H}\geq Z_q^{\mathrm L}$, we combine
\cref{lem:capped-q-heavy-dual-fitting} with the bound
$\opt_{s,q}\leq\opt_{1,q}/s$ for $s=1/(8q)$. When
$Z_q^{\mathrm L}\geq Z_q^{\mathrm H}$, we combine
\cref{lem:capped-q-light-dual-fitting} with this bound for $s=B/(4qD)$.
This gives the respective bounds $\alg_q/\opt_{1,q}\leq\mathrm{cr}_1$ and
$\alg_q/\opt_{1,q}\leq\mathrm{cr}_2$, where
\begin{align*}
    \mathrm{cr}_1
    &:=8q\biggl(1+(8B)^{1/q}\biggr)
      \leq72qB^{1/q},\\
    \mathrm{cr}_2
    &:=\frac{4qD}{B}\biggl(1+4^{1/q}\biggr)
      \leq\frac{20qD}{B}.
\end{align*}
Here we used $8^{1/q}\leq8$, $4^{1/q}\leq4$, and $B^{1/q}\geq1$.
At least one of the two cases holds, so
\[
    \frac{\alg_q}{\opt_{1,q}}
    \leq\max\{\mathrm{cr}_1,\mathrm{cr}_2\}
    \leq72q\max\biggl\{B^{1/q},\frac{D}{B}\biggr\}.
\]
We choose $B:=D^{q/(q+1)}$, which lies in $[1,D]$. This choice makes the
two terms in the maximum equal, since
\[
    B^{1/q}=\frac{D}{B}=D^{1/(q+1)}.
\]
We obtain $\alg_q\leq72qD^{1/(q+1)}\opt_{1,q}$ when all jobs are released
at time zero.

Finally, by \cref{lem:full-drf-dominates-one-shot}, full DRF completes each
job no later than one-shot DRF on the same instance. Its $\ell_q$ norm is
therefore no larger, so the same guarantee holds for full DRF.
\end{proof}

\subsection{An Upper Bound Independent of \texorpdfstring{$q$}{q}}
\label{sec:lq-dimension-upper-bound}

We now give an upper bound that is linear in $D$ and does not grow with
$q$. It improves upon \cref{thm:lq-q-upper-bound} for large $q$, in
particular when $q\geq D$. We reuse the cap separation from
\cref{lem:capped-q-effective-decomposition} and bound the time jobs spend
below their rate caps. The idea is that a job running below its cap
receives at least the rate of every other job, while the total rate of all
jobs is at least one. Hence, every job reaches its cap no later than it
would complete under processor sharing on a single unit-speed machine. In
contrast, even an optimal schedule can process at most $D$ units of work per
unit of time.

\begin{theorem}
\label{thm:lq-dimension-upper-bound}
For every integer $D\geq1$ and real $q\geq1$, one-shot DRF and full DRF
have a competitive ratio in $O(D)$ for minimizing the $\ell_q$ norm of
completion times in multidimensional scheduling with $D$ resources and zero-release times.
\end{theorem}

\begin{proof}
Assume that all jobs are released at time zero, and consider
one-shot DRF.
Let $\tau=(\tau_j)_{j=1}^n$ be as in
\cref{lem:capped-q-effective-decomposition}. By that lemma with $s=1$,
it suffices to show $\lVert\tau\rVert_q\leq2D \cdot \opt_{1,q}$.

Fix a job $j$. For $t<\tau_j$, job $j$ receives rate
$x_j(t)=\lambda(t)<u_j$. Every other unfinished job receives at most
this rate, and completed jobs receive rate zero. Since $c(t)>0$,
\eqref{eq:capped-q-effective-rate} and
\eqref{eq:capped-q-effective-congestion} give
\[
    \sum_{i=1}^n x_i(t)=\frac{m(t)}{c(t)}\geq1.
\]
The inequality holds because $c(t)=\max_d c_d(t)\leq m(t)$, as $a_{id}\leq1$.
Thus the schedule processes at least one unit of work per unit of time
throughout $[0,\tau_j)$. At the same time, every job $i$ has received no
more processing than job $j$, and no job receives more than its own
processing requirement. More precisely,
\begin{align*}
    \int_0^{\tau_j}x_i(t)\,\mathrm{d}t \leq p_i \quad \text{and} \quad
    \int_0^{\tau_j}x_i(t)\,\mathrm{d}t
    \leq\int_0^{\tau_j}x_j(t)\,\mathrm{d}t \leq p_j.
\end{align*}
Consequently,
\begin{equation}
\label{eq:lq-jobwise-dimension-bound}
    \tau_j
    \leq\int_0^{\tau_j}\sum_{i=1}^n x_i(t)\,\mathrm{d}t
    \leq\sum_{i=1}^n\min\{p_i,p_j\}
    =:P_j.
\end{equation}

Now relabel the jobs so that $p_1\leq\cdots\leq p_n$, and let
$S_j:=\sum_{i=1}^jp_i$. Then $P_j=S_j+(n-j)p_j$. Moreover, it holds that
\[
    P_{j+1}-P_j=(n-j)(p_{j+1}-p_j)\geq0
\]
for every $j \in [n-1]$,
so the sequence $(P_j)_{j=1}^n$ is nondecreasing. Let
$P:=(P_j)_{j=1}^n$ and $S:=(S_j)_{j=1}^n$ denote the corresponding
vectors. We have
\begin{align*}
    \sum_{j=1}^n P_j^q
    &=\sum_{j=1}^n S_jP_j^{q-1}
      +\sum_{i=1}^n(n-i)p_iP_i^{q-1}
    =\sum_{j=1}^n S_jP_j^{q-1}
      +\sum_{j=1}^n\sum_{i=1}^{j-1}p_iP_i^{q-1}\\
    &\leq\sum_{j=1}^n S_jP_j^{q-1}
      +\sum_{j=1}^n\sum_{i=1}^{j-1}p_iP_j^{q-1}
    =\sum_{j=1}^n(2S_j-p_j)P_j^{q-1}
     \leq2\sum_{j=1}^n S_jP_j^{q-1}.
\end{align*}
The second equality writes $p_iP_i^{q-1}$ once for each index $j>i$, hence
exactly $n-i$ times. 
For the first inequality, note that if $i<j$, then
$P_i\leq P_j$, and therefore $p_iP_i^{q-1}\leq p_iP_j^{q-1}$ because
$q\geq1$. The final equality uses
$\sum_{i=1}^{j-1}p_i=S_j-p_j$.
For $q=1$, this already says $\lVert P\rVert_1\leq2\lVert S\rVert_1$.
For $q>1$, Hölder's inequality\footnote{For any
$a,b\in\mathbb{R}^n$ and exponents $r,s>1$ with $1/r+1/s=1$,
Hölder's inequality states that
$\sum_{j=1}^n|a_jb_j|\leq
(\sum_{j=1}^n|a_j|^r)^{1/r}
(\sum_{j=1}^n|b_j|^s)^{1/s}$.}
with exponents $q$ and $q/(q-1)$ gives
\[
    \lVert P\rVert_q^q
    \leq2\sum_{j=1}^n S_jP_j^{q-1}
    \leq2\lVert S\rVert_q \cdot \lVert P\rVert_q^{q-1}.
\]
Dividing by $\lVert P\rVert_q^{q-1}$ shows that
$\lVert P\rVert_q\leq2\lVert S\rVert_q$ for every $q\geq1$.

Let $C_j^*$ and $y_j^*(t)$ be the completion time and rate of job $j$ in an
optimal unit-speed schedule, and define $C^*:=(C_j^*)_{j=1}^n$. Since
$\lVert a_j\rVert_\infty=1$ and demands are nonnegative, we have
$\sum_{d=1}^D a_{jd}\geq1$ for every job $j$, so
\[
    \sum_{j=1}^n y_j^*(t)
    \leq\sum_{d=1}^D\sum_{j=1}^n a_{jd}y_j^*(t)
    \leq D.
\]
Let $C_{(j)}^*$ denote the $j$th smallest optimal completion time.
By time $C_{(j)}^*$, at least $j$ jobs have completed. Their total processing
requirement is at least $S_j$, the sum of the $j$ smallest processing
requirements. Hence $S_j\leq D \cdot C_{(j)}^*$. Since the norm is symmetric, $\lVert S\rVert_q\leq D \cdot \lVert C^*\rVert_q$.
Combining this with \eqref{eq:lq-jobwise-dimension-bound} and
\cref{lem:capped-q-effective-decomposition} at $s=1$ gives
\begin{align*}
    \alg_q
    \leq\opt_{1,q}+\lVert\tau\rVert_q
    \leq\opt_{1,q}+\lVert P\rVert_q
    \leq\opt_{1,q}+2\lVert S\rVert_q
    \leq(2D+1)\opt_{1,q}.
\end{align*}

By \cref{lem:full-drf-dominates-one-shot}, full DRF completes every job no
later than one-shot DRF, so the same bound holds for full DRF.
\end{proof}

\section{Conclusion}
\label{sec:conclusion}

We establish asymptotically tight worst-case completion-time guarantees
for DRF in multidimensional nonclairvoyant scheduling. With $D$ resources,
DRF is $\Theta(\sqrt D)$-competitive for total completion time and
$\Theta(D^{1/(q+1)})$-competitive for the $\ell_q$ norm of completion
times for each fixed real $q\geq1$. These guarantees hold with arbitrary
release times and without advance knowledge of job sizes. We also obtain
sharper bounds for total completion time with two and three resources.

The analysis combines lower bounds based on shared bottlenecks with
those based on aggregate processing capacity. Cauchy--Schwarz combines
these two perspectives for total completion time, while LP dual fitting
extends the approach to general norms. The resulting bounds quantify
how DRF's instantaneous fairness translates into completion-time
performance, and identify the dependence on the number of resources.
Determining the optimal competitive constants, especially for small
numbers of resources, remains an interesting direction.

\section*{Methods and AI Use}

The authors initiated this study with the binary-demand case and developed
complete proofs; earlier attempts to obtain these proofs using AI were
unsuccessful. We used OpenAI Codex and Anthropic Claude Code to assist with drafting and revising
text, searching and checking references, preparing illustrations, and
checking mathematical arguments. AI assistance was also used to generalize
some arguments, including extending them to incorporate job-specific rate
caps. The authors verified all content, including the proofs and references,
and take full responsibility for its correctness.
 
\section*{Acknowledgments}
AL was supported by the Deutsche Forschungsgemeinschaft (DFG, German Research Foundation) under Germany’s Excellence Strategy -- The Berlin Mathematics Research Center MATH$^+$ (EXC-2046/1, EXC-2046/2, project ID: 390685689).
MB and SI were supported in part by NSF grants CCF-2423106, CCF-2537126, and CCF-2535599.

\printbibliography

\appendix

\section{Omitted Details from \texorpdfstring{\Cref{sec:model-and-algorithms}}{Model and Algorithms}}
\label{app:model}

\subsection{Full DRF Dominates One-Shot DRF}

\lemFullDrfDominatesOneShot*

\begin{proof}
Run both algorithms on the same instance, with identical release times,
processing requirements, demands, and caps. For
$S\in\{\mathrm{DRF},\mathrm{OS}\}$, let $A_S(t)$ be its active set and let
\[
    e_j^S(t):=\int_0^t x_j^S(s)\,\mathrm{d}s
\]
be the work it has processed on job $j$ by time $t$, taking the rate to be
zero before release and after completion. Thus $0\leq e_j^S(t)\leq p_j$.
We prove the invariant
\[
    e_j^{\mathrm{DRF}}(t)\geq e_j^{\mathrm{OS}}(t)
\]
for every job $j$ and time $t\geq0$.
This invariant implies $A_{\mathrm{DRF}}(t)\subseteq A_{\mathrm{OS}}(t)$, because
if $j$ is released and unfinished under full DRF, then
$e_j^{\mathrm{OS}}(t)\leq e_j^{\mathrm{DRF}}(t)<p_j$, so it is also
unfinished under one-shot DRF.

We induct over the release times and the completion times in either
schedule. Initially, both schedules have processed zero work. Suppose the
invariant holds at an event time $t_0$, after all releases and completions
at that time have been accounted for. Until the next event time $t_1$, both
active sets remain unchanged. For every $j\in A_{\mathrm{DRF}}(t_0)$ and
$t\in[t_0,t_1)$, we therefore have
\[
    x_j^{\mathrm{DRF}}(t)
    =x_j^{\mathrm{DRF}}(A_{\mathrm{DRF}}(t_0))
    \geq x_j^{\mathrm{OS}}(A_{\mathrm{DRF}}(t_0))
    \geq x_j^{\mathrm{OS}}(A_{\mathrm{OS}}(t_0))
    =x_j^{\mathrm{OS}}(t).
\]
The first inequality holds because full DRF only increases rates after the
initial round; the second follows from
\cref{obs:one-shot-monotonicity} and the active-set inclusion at $t_0$.
Integrating this inequality from $t_0$ to $t$ preserves the work invariant
for these jobs. Every other job either has not yet been released, so both
work values are zero, or has already completed under full DRF, so
$e_j^{\mathrm{DRF}}(t)=p_j\geq e_j^{\mathrm{OS}}(t)$.

The work functions are continuous, so the invariant also holds at $t_1$.
In particular, if a job completes under one-shot DRF at $t_1$, then full
DRF has already processed at least $p_j$ units of that job and has completed
it by $t_1$. Any newly released job starts with zero processed work in both
schedules. Hence the invariant and the active-set inclusion survive all
events at $t_1$, including simultaneous releases and completions. This
completes the induction. At $t=C_j^{\mathrm{OS}}$, the invariant gives
$e_j^{\mathrm{DRF}}(t)\geq e_j^{\mathrm{OS}}(t)=p_j$, proving
$C_j^{\mathrm{DRF}}\leq C_j^{\mathrm{OS}}$ for every job $j$.
\end{proof}

\subsection{Equivalence with the Unnormalized Formulation}
\label{app:normalization}

The original DRF paper \cite{ghodsi2011drf} uses unnormalized demand vectors.
In the corresponding scheduling formulation, job $j$ has a processing
requirement $\widetilde p_j>0$, a nonzero demand vector
$\widetilde a_j=(\widetilde a_{j1},\ldots,\widetilde a_{jD})
\in\mathbb{R}_{\geq0}^D$ whose infinity norm may be arbitrary, and a unit
rate cap. Here, $\widetilde a_{jd}$ is the amount of resource $d$ that job $j$
uses when it runs at full speed. A schedule
$\widetilde x$ satisfies $0\leq\widetilde x_j(t)\leq1$, together with the
corresponding resource-capacity and completion constraints.
Original DRF applies progressive filling to the jobs' dominant shares rather
than to their raw rates. It freezes jobs that use a tight resource and
continues with the remaining jobs.

\Cref{alg:full-drf} instead uses normalized demand vectors and
job-dependent rate caps. The two formulations are equivalent, and converting
between them only changes the units in which demands, rates, and processing
requirements are expressed. Under this conversion, original DRF and
\cref{alg:full-drf} give every job the same amount of every resource, and
every job has the same completion time in both formulations. We use the
normalized formulation because $x_j$ is then exactly job $j$'s dominant
share, which makes both progressive filling and our analysis easier to
express. If the original demand vectors already have unit infinity norm, then
$u_j=1$ for every job and \cref{alg:full-drf} is exactly the original DRF
allocation.

\begin{lemma}[Equivalence of the two formulations]
\label{lem:normalization-equivalence}
There is a one-to-one correspondence between instances of the normalized
formulation with job-dependent rate caps and instances of the unnormalized
formulation with unit rate caps. For every pair of corresponding instances,
there is a one-to-one correspondence between their feasible schedules that
preserves every job's completion time.
\end{lemma}

\begin{proof}
First, consider an instance of the unnormalized formulation. For every job
$j$, let $\Delta_j:=\lVert\widetilde a_j\rVert_\infty>0$, and construct the
corresponding normalized job $j$ by setting
\begin{align*}
	a_j &:= \frac{\widetilde a_j}{\Delta_j}, \\ 
	u_j &:= \Delta_j, \\
	p_j &:= \Delta_j\widetilde p_j.
\end{align*}
The release time $r_j$ is unchanged. By construction, we have
$\lVert a_j\rVert_\infty=1$.

Conversely, given an instance of the normalized formulation, define the
corresponding unnormalized job by
\begin{align*}
	\widetilde a_j &:= u_ja_j, \\
	\widetilde p_j &:= \frac{p_j}{u_j}.
\end{align*}
The release time $r_j$ is unchanged. Since $\lVert a_j\rVert_\infty=1$, we have
$\lVert\widetilde a_j\rVert_\infty=u_j$. Hence the two instance
transformations are inverses.

It remains to relate schedules on corresponding instances. Given a schedule
$\widetilde x$ for the unnormalized instance, define
\[
    x_j(t):=u_j\widetilde x_j(t)
    \qquad\text{for every job $j$ and time $t$}.
\]
Conversely, recover $\widetilde x$ from $x$ by setting
$\widetilde x_j(t):=x_j(t)/u_j$. These maps are inverses. Furthermore,
$\widetilde x_j(t)\in[0,1]$ if and only if $x_j(t)\in[0,u_j]$, and
\[
    a_{jd}x_j(t)=\widetilde a_{jd}\widetilde x_j(t),
\]
so the two schedules have identical resource consumption at every time.
Finally, because $p_j=u_j\widetilde p_j$,
\[
    \int_{r_j}^t x_j(s)\,\mathrm{d}s\geq p_j
    \quad\Longleftrightarrow\quad
    \int_{r_j}^t \widetilde x_j(s)\,\mathrm{d}s\geq\widetilde p_j.
\]
Thus the schedule correspondence preserves feasibility and every job's
completion time.
\end{proof}

The correspondence also preserves the allocation selected by DRF. Indeed,
consider an unnormalized job with demand vector $\widetilde a_j$, and let
$\Delta_j=\lVert\widetilde a_j\rVert_\infty$. If $\widetilde x_j$ is its
rate, then its dominant share is $\Delta_j\widetilde x_j$, which is exactly
the corresponding normalized rate $x_j=u_j\widetilde x_j$. Therefore,
increasing the dominant shares of all unfrozen jobs at the same speed in the
unnormalized instance is equivalent to increasing their normalized rates at
the same speed. Moreover, we have
$a_{jd}x_j=\widetilde a_{jd}\widetilde x_j$. So although the scalar rates are
expressed in different units, every job receives exactly the same amount of
each resource in the two instances. Consequently, the same resources become
tight and the same jobs are frozen in every round. Thus, if
$\widetilde x^{\mathrm{DRF}}$ and $x^{\mathrm{DRF}}$ are the DRF allocations
for corresponding unnormalized and normalized instances, respectively, then
$x_j^{\mathrm{DRF}}=u_j\widetilde x_j^{\mathrm{DRF}}$ for every job $j$.
Together with \cref{lem:normalization-equivalence}, this shows that original
DRF and \cref{alg:full-drf} give every job the same resource
allocation at every time and the same completion time.

\section{Low-Dimensional Bounds}
\label{sec:low-dimensional-bounds}

We now sharpen \cref{thm:drf-completion-time} for two and three resources.
These cases are of particular practical interest, since cluster jobs
typically request only a few resources such as CPU and memory. For $D=2$ and
$D=3$,
\cref{thm:drf-completion-time} gives competitive ratios of
$4\sqrt2+1\approx6.66$ and $4\sqrt3+1\approx7.93$. We improve these to
$2+\sqrt3\approx3.73$ and $2+\sqrt7\approx4.65$, respectively.

We retain the zero-release assumption and the notation from
\cref{sec:completion-time}, in particular the phase notation from
\cref{subsec:bottleneck-bound}.
Recall that
$z_{j,h}$ is the effective mass of job $j$ in phase $h$, $l_h$ is the
uncapped service amount of that phase, $m_h=\sum_{j=1}^nz_{j,h}$, and $c_h$ is the congestion.

The general analysis loses a factor of $\sqrt D$ because
\cref{lem:effective-mass-bound} only uses the total capacity of all
resources. For few resources, we can afford to derive lower bounds on $\opt$
from each resource separately. To this end, we view each resource as a
unit-speed single machine. On resource $d$, job $j$ requires
$a_{jd}p_j$ units of processing. Any feasible multidimensional schedule
induces a feasible schedule on this machine in which each job finishes
no later than in the original schedule. However, simply adding the
unweighted completion costs of these projected schedules could count
the same job's completion cost several times.

To avoid this double counting, we use auxiliary \emph{weighted}
single-machine instances. In such an instance, job $j$ has a processing
requirement $\bar p_j$ and a nonnegative weight $v_j$, and the objective is
$\sum_{j=1}^nv_jC_j$, called the \emph{total weighted completion time}.
The auxiliary instances also have all release times zero. In our
applications, the weights
assigned to a job across the resource projections sum to one. They
therefore split its original completion cost among the resources,
rather than counting that cost in full for each resource. The sum of
the projected optimal weighted costs is then a lower bound on $\opt$.
These weights are only a tool for the analysis. The multidimensional
problem and the DRF algorithm remain unweighted.
We choose the weights according to the resource demands so that the
resulting lower bounds on $\opt$ can be expressed in terms of the
resource loads that determine one-shot DRF's processing rates.

To apply the next lemma to resource $d$, we choose a weight
$v_j\leq a_{jd}$ and set $q_{j,h}:=v_jz_{j,h}$. During phase $h$,
one-shot DRF performs $l_hz_{j,h}$ units of processing on job $j$,
using $a_{jd}l_hz_{j,h}$ units of resource $d$. We count only
$l_hq_{j,h}=v_jl_hz_{j,h}$ units of this resource usage.
Because $v_j\leq a_{jd}$, the amount counted is no larger than the
amount actually used. Its sum over the phases is therefore at most
the projected processing requirement $\bar p_j=a_{jd}p_j$.
Also, $q_{j,h}\leq v_j$ because $z_{j,h}\leq1$.
The following lemma uses precisely these two inequalities to lower-bound
the optimal weighted completion time.

\begin{restatable}{lemma}{lemEffectiveProfileLowerBound}
\label{lem:effective-profile-lower-bound}
Consider $n$ jobs with processing requirements $\bar p_j\geq0$ and weights $v_j\geq0$
on a unit-speed single machine. Suppose that, for each job $j$ and phase
$h\in[r]$, we are given a value $q_{j,h}$ satisfying
$0\leq q_{j,h}\leq v_j$ and $\sum_{h=1}^r l_hq_{j,h}\leq\bar p_j$.
If $Q_h:=\sum_{j=1}^nq_{j,h}$, then
\[
    \opt_v\geq\frac12\sum_{h=1}^r l_hQ_h^2 \ ,
\]
where $\opt_v$ is the optimal weighted completion time on the single machine.
\end{restatable}

\begin{proof}
On a single machine with all release times zero, Smith's rule is optimal for
total weighted completion time. It schedules the jobs in nonincreasing order
of $v_j/\bar p_j$. The completion time of a job consists of its own
processing requirement and those of all jobs scheduled before it. Hence,
the optimal value is
\[
    \opt_v=\sum_{j=1}^nv_j\bar p_j
    +\sum_{j=1}^{n-1}\sum_{k=j+1}^n\min\{v_j\bar p_k,v_k\bar p_j\}.
\]

We compare this expression with the individual-job and pairwise terms in
$\frac12\sum_{h=1}^r l_hQ_h^2$. For every job $j$, the assumptions on $q_{j,h}$ give
\[
    \sum_{h=1}^r l_hq_{j,h}^2
    \leq v_j\sum_{h=1}^r l_hq_{j,h}
    \leq v_j\bar p_j.
\]
For two distinct jobs $j,k$, they similarly give
\[
    \sum_{h=1}^r l_hq_{j,h}q_{k,h}
    \leq v_k\sum_{h=1}^r l_hq_{j,h}
    \leq v_k\bar p_j.
\]
Interchanging $j$ and $k$ also bounds this sum by $v_j\bar p_k$, so it is at
most $\min\{v_j\bar p_k,v_k\bar p_j\}$. Expanding $Q_h^2$ now yields
\begin{align*}
    \frac12\sum_{h=1}^r l_hQ_h^2
    &=\frac12\sum_{j=1}^n\sum_{h=1}^r l_hq_{j,h}^2
      +\sum_{j=1}^{n-1}\sum_{k=j+1}^n\sum_{h=1}^r l_hq_{j,h}q_{k,h}\\
    &\leq\frac12\sum_{j=1}^nv_j\bar p_j
      +\sum_{j=1}^{n-1}\sum_{k=j+1}^n\min\{v_j\bar p_k,v_k\bar p_j\}\\
    &\leq\opt_v.
\end{align*}
This completes the proof of the lemma.
\end{proof}

We use this lemma to bound $Z_1=\sum_{h=1}^r l_hm_hc_h$. For each of the two
cases below, we first obtain a lower bound on $\opt$ from weighted
resource projections, and then compare it with $m_hc_h$ phase by phase.
The decomposition $\alg\leq\opt+Z_1$ from
\cref{lem:effective-time-decomposition} then gives the competitive ratio.

\subsection{Two Resources}

For every phase $h$, let $A_h:=\sum_{j=1}^na_{j1}z_{j,h}$ and
$B_h:=\sum_{j=1}^na_{j2}z_{j,h}$ be the effective loads of the two resources.
By \eqref{eq:capped-q-effective-congestion}, we have
$c_h=\max\{A_h,B_h\}$. Demand normalization also gives
$A_h+B_h\geq m_h$, while $a_{jd}\leq1$ implies $A_h \le m_h$ and $B_h\leq m_h$.

\begin{restatable}[Two-resource projection bound]{lemma}{lemTwoResourceProjection}
\label{lem:two-resource-effective-projection}
For $D=2$ resources, it holds that
\[
    \opt\geq\frac14\sum_{h=1}^r l_h
    \left(
        A_h^2+(m_h-A_h)^2+(m_h-B_h)^2+B_h^2
    \right).
\]
\end{restatable}

\begin{proof}
Fix an optimal multidimensional schedule with completion times $C_j^*$.
We first split job $j$'s completion cost between the two resources using
weights $a_{j1}$ and $1-a_{j1}$.

Project the schedule onto resource $1$, giving job $j$ processing
requirement $a_{j1}p_j$ and weight $a_{j1}$. The projected schedule is
feasible on a unit-speed single machine and has weighted completion time
at most $\sum_{j=1}^na_{j1}C_j^*$. On the other hand,
\cref{lem:effective-profile-lower-bound} with
$q_{j,h}:=a_{j1}z_{j,h}$ lower-bounds the optimum of this single-machine
instance. Hence
\[
    \sum_{j=1}^na_{j1}C_j^*\geq\frac12\sum_{h=1}^r l_hA_h^2.
\]
For resource $2$, give job $j$ processing requirement $a_{j2}p_j$ and weight
$1-a_{j1}$, and set $q_{j,h}:=(1-a_{j1})z_{j,h}$. Normalization gives
$a_{j1}+a_{j2}\geq1$, so $1-a_{j1}\leq a_{j2}$ and these values again
satisfy the lemma's assumptions. Since
$Q_h=\sum_{j=1}^n(1-a_{j1})z_{j,h}=m_h-A_h$, the same argument gives
\[
    \sum_{j=1}^n(1-a_{j1})C_j^*\geq\frac12\sum_{h=1}^r l_h(m_h-A_h)^2.
\]
Since the two weights of each job sum to one, adding these inequalities
counts each job's completion cost exactly once. This gives
\[
    \opt=\sum_{j=1}^nC_j^*\geq\frac12\sum_{h=1}^r l_h
    \left(A_h^2+(m_h-A_h)^2\right).
\]
Interchanging the roles of the resources, using weights $1-a_{j2}$ and
$a_{j2}$ instead, yields
\[
    \opt\geq\frac12\sum_{h=1}^r l_h
    \left((m_h-B_h)^2+B_h^2\right).
\]
Averaging the two inequalities proves the lemma.
\end{proof}

\begin{restatable}{lemma}{lemTwoResourcePointwise}
\label{lem:two-resource-pointwise}
For all $x,y\in[0,1]$ with $x+y\geq1$, it holds that
\[
    \max\{x,y\}
    \leq
    \frac{1+\sqrt3}{4}
    \left(x^2+(1-x)^2+(1-y)^2+y^2\right).
\]
\end{restatable}

\begin{proof}
By symmetry, assume $x\geq y$. Since $x+y\geq1$, we have $x\geq1/2$.
For fixed $x$, the term $(1-y)^2+y^2=2(y-1/2)^2+1/2$ is minimized at
$y=1/2$. This choice is feasible because it satisfies both $y\leq x$ and
$x+y\geq1$. Therefore,
\[
    \frac{4\max\{x,y\}}{x^2+(1-x)^2+(1-y)^2+y^2}
    \leq\frac{4x}{2x^2-2x+3/2}.
\]
The derivative of the right-hand side has the sign of $3/2-2x^2$.
Thus its maximum on $[1/2,1]$ occurs at $x=\sqrt3/2$ and equals
$1+\sqrt3$, proving the claim.
\end{proof}

\begin{theorem}[Two resources]
\label{thm:drf-two-resources}
For multidimensional scheduling with $D=2$ and zero-release times,
one-shot DRF and full DRF are $(2+\sqrt3)$-competitive for minimizing
total completion time.
\end{theorem}

\begin{proof}
For every phase with $m_h>0$, set $x:=A_h/m_h$ and $y:=B_h/m_h$.
These normalized loads lie in $[0,1]$ and satisfy $x+y\geq1$, so we may
apply \cref{lem:two-resource-pointwise}. Multiplying its inequality by
$m_h^2$ and using $c_h=\max\{A_h,B_h\}$ gives
\[
    m_hc_h
    \leq
    \frac{1+\sqrt3}{4}
    \left(
        A_h^2+(m_h-A_h)^2+(m_h-B_h)^2+B_h^2
    \right).
\]
Multiplying by $l_h$, summing over the phases, and applying
\cref{lem:two-resource-effective-projection}, we obtain
\[
    Z_1=\sum_{h=1}^r l_hm_hc_h\leq(1+\sqrt3)\opt.
\]
The effective-time decomposition in \cref{lem:effective-time-decomposition}
now gives
\[
    \alg\leq\opt+Z_1\leq(2+\sqrt3)\opt.
\]
This proves the bound for one-shot DRF. Full DRF satisfies the same bound
by \cref{lem:full-drf-dominates-one-shot}.
\end{proof}

\subsection{Three Resources}

For each phase $h$ and resource $d\in[3]$, define the effective load
$A_{d,h}:=\sum_{j=1}^na_{jd}z_{j,h}$. As before, we have
$c_h=\max_dA_{d,h}$, $0\leq A_{d,h}\leq m_h$, and
$\sum_{d=1}^3A_{d,h}\geq m_h$. We use the same approach as for two resources,
but now split the weight remaining after one projection between the two
other resources.

\begin{restatable}[Three-resource projection bound]{lemma}{lemThreeResourceProjection}
\label{lem:three-resource-effective-projection}
For $D=3$ resources, it holds that
\[
    \opt
    \geq
    \frac16\sum_{h=1}^r l_h
    \sum_{d=1}^3
    \left(
        A_{d,h}^2+\frac{(m_h-A_{d,h})^2}{2}
    \right).
\]
\end{restatable}

\begin{proof}
Fix a resource $d\in[3]$. We assign weight $a_{jd}$ to job $j$ in
projection $d$, and distribute the remaining weight $1-a_{jd}$ between
the other two projections. Each assigned weight must be at most the
corresponding demand. Such a split is possible because normalization gives
$\sum_{\substack{d'=1\\d'\neq d}}^3a_{j d'}\geq1-a_{jd}$. Denote the resulting weights by
$\gamma_{j1},\gamma_{j2},\gamma_{j3}$, so
$0\leq\gamma_{j d'}\leq a_{j d'}$, $\sum_{d'=1}^3\gamma_{j d'}=1$, and $\gamma_{jd}=a_{jd}$.

Project an optimal multidimensional schedule onto each resource $d'$. In
projection $d'$, give job $j$ processing requirement $a_{j d'}p_j$ and weight
$\gamma_{j d'}$, and set $q_{j,h}:=\gamma_{j d'}z_{j,h}$. These values
satisfy the assumptions of \cref{lem:effective-profile-lower-bound}
because $\gamma_{j d'}\leq a_{j d'}$. As in the two-resource case, the
weights sum to one for each job, so the sum of the three projected optima
is at most $\opt$. Set $Q_{d',h}:=\sum_{j=1}^n\gamma_{j d'}z_{j,h}$.
Summing the three applications of
\cref{lem:effective-profile-lower-bound} gives
\[
    \opt\geq\frac12\sum_{h=1}^r l_h\sum_{d'=1}^3Q_{d',h}^2.
\]

By construction, $Q_{d,h}=A_{d,h}$ and
$\sum_{\substack{d'=1\\d'\neq d}}^3Q_{d',h}=m_h-A_{d,h}$. For a fixed sum, the sum of the
squares of the two remaining loads is smallest when they are equal.
Formally, Cauchy--Schwarz gives
\[
    \sum_{\substack{d'=1\\d'\neq d}}^3Q_{d',h}^2
    \geq\frac12\biggl(\sum_{\substack{d'=1\\d'\neq d}}^3Q_{d',h}\biggr)^2
    =\frac{(m_h-A_{d,h})^2}{2}.
\]
Consequently,
\[
    \opt
    \geq
    \frac12\sum_{h=1}^r l_h
    \left(
        A_{d,h}^2+\frac{(m_h-A_{d,h})^2}{2}
    \right).
\]
Averaging this inequality over the three choices of $d$ proves the lemma.
\end{proof}

\begin{restatable}{lemma}{lemThreeResourcePointwise}
\label{lem:three-resource-pointwise}
For all $x_1,x_2,x_3\in[0,1]$ with $\sum_{d=1}^3x_d\geq1$, it holds that
\[
    \max_d x_d
    \leq
    \frac{1+\sqrt7}{6}
    \sum_{d=1}^3
    \left(x_d^2+\frac{(1-x_d)^2}{2}\right).
\]
\end{restatable}

\begin{proof}
Let $g:=\max_d x_d$. Since $\sum_{d=1}^3x_d\geq1$, we have $g\in[1/3,1]$.
The identity
\[
    x^2+\frac{(1-x)^2}{2}
    =\frac13+\frac32\left(x-\frac13\right)^2
\]
shows that each summand on the right-hand side is minimized at $x=1/3$.
For fixed $g$, we therefore minimize their sum by setting the other two
coordinates to $1/3$. This choice is feasible because both coordinates
are at most $g$ and $g+2/3\geq1$. The sum is then
$g^2+(1-g)^2/2+2/3=(9g^2-6g+7)/6$, so
\[
    \frac{6\max_d x_d}
         {\sum_{d=1}^3\left(x_d^2+(1-x_d)^2/2\right)}
    \leq\frac{36g}{9g^2-6g+7}.
\]
The derivative of the right-hand side has the sign of $7-9g^2$.
Thus its maximum on $[1/3,1]$ occurs at $g=\sqrt7/3$ and equals
$1+\sqrt7$, proving the claim.
\end{proof}

\begin{theorem}[Three resources]
\label{thm:drf-three-resources}
For multidimensional scheduling with $D=3$ and zero-release times,
one-shot DRF and full DRF are $(2+\sqrt7)$-competitive for minimizing
total completion time.
\end{theorem}

\begin{proof}
For every phase with $m_h>0$, set $x_d:=A_{d,h}/m_h$. These normalized
loads lie in $[0,1]$ and satisfy $\sum_{d=1}^3x_d\geq1$. Apply
\cref{lem:three-resource-pointwise}, multiply by $m_h^2$, and use
$c_h=\max_dA_{d,h}$ to obtain
\[
    m_hc_h
    \leq
    \frac{1+\sqrt7}{6}
    \sum_{d=1}^3
    \left(
        A_{d,h}^2+\frac{(m_h-A_{d,h})^2}{2}
    \right).
\]
Multiplying by $l_h$, summing over phases, and applying
\cref{lem:three-resource-effective-projection} gives
\[
    Z_1=\sum_{h=1}^r l_hm_hc_h\leq(1+\sqrt7)\opt.
\]
By \cref{lem:effective-time-decomposition},
\[
    \alg\leq\opt+Z_1\leq(2+\sqrt7)\opt.
\]
The same bound holds for full DRF by
\cref{lem:full-drf-dominates-one-shot}.
\end{proof}

\section{Omitted Proofs from \texorpdfstring{\Cref{sec:lq-general-case}}{lq Norms}}
\label{app:lq-general-case}

\subsection{Omitted Proofs for the Upper Bound Depending on \texorpdfstring{$q$}{q}}

\lemCappedQEffectiveDecomposition*

\begin{proof}
Because $\lambda(t)$ is nondecreasing, a job that reaches its cap stays at
that cap until it finishes. In particular, $\tau_j=0$ when the job starts
at its cap.
Before time $\tau_j$, the cap does not restrict the job's rate, so $z_j(t)=1$.
After time $\tau_j$, any remaining work takes at most $p_j/u_j$ time. Hence
\[
    Z_{j,q}\geq\int_0^{\tau_j}\omega(t)\,\mathrm{d}t=\tau_j^q
\]
and
\[
    C_j\leq\tau_j+\frac{p_j}{u_j}
        \leq Z_{j,q}^{1/q}+ \frac{p_j}{u_j}.
\]
Thus, we have in total
\begin{align*}
    \alg_q
    =\biggl(\sum_{j=1}^n C_j^q\biggr)^{1/q}
    &\leq\biggl(\sum_{j=1}^n\biggl(\frac{p_j}{u_j}+\tau_j\biggr)^q\biggr)^{1/q}\\
    &\leq\biggl(\sum_{j=1}^n \biggl(\frac{p_j}{u_j}\biggr)^q\biggr)^{1/q}
       +\biggl(\sum_{j=1}^n\tau_j^q\biggr)^{1/q}
    =\mathrm{CAP}_q+\lVert\tau\rVert_q.
\end{align*}
The second inequality is the triangle inequality for the $\ell_q$ norm.
The last equality uses the definitions of $\mathrm{CAP}_q$ and $\tau$.
Finally, applying \eqref{eq:lb-q-opt-ineq} and
$\tau_j^q\leq Z_{j,q}$ gives, for every $s\in(0,1]$,
\[
    \alg_q\leq\opt_{s,q}+\lVert\tau\rVert_q
    \leq\opt_{s,q}+Z_q^{1/q}.
\]
This concludes the proof of the lemma.
\end{proof}

\lemCappedQPrimalRelaxation*

\begin{proof}
Consider an optimal speed-$s$ schedule with completion times $C_j^*$.
Let $y_j^*(t)$ be its rate for job $j$ until completion, and set $y_j^*(t)=0$
for $t\geq C_j^*$. These rates satisfy the primal's resource constraints
and rate caps because the schedule is feasible. Moreover, each job
receives exactly its processing requirement by completion, so
\[
    \frac1{p_j}\int_0^\infty y_j^*(t)\,\mathrm{d}t
    =\frac1{p_j}\int_0^{C_j^*}y_j^*(t)\,\mathrm{d}t=1.
\]
Thus $y^*$ is a feasible primal solution. Since $t^q\leq(C_j^*)^q$ for
$0\leq t\leq C_j^*$, its objective value satisfies
\begin{align*}
    \sum_{j=1}^n\frac1{p_j}\int_0^\infty t^qy_j^*(t)\,\mathrm{d}t
    &\leq\sum_{j=1}^n\frac{(C_j^*)^q}{p_j}
        \int_0^{C_j^*}y_j^*(t)\,\mathrm{d}t=\sum_{j=1}^n(C_j^*)^q
     =\opt_{s,q}^q.
\end{align*}
The primal's optimal value is no larger than the value of this feasible
solution, which proves the claim.
\end{proof}

\lemCappedQDualBound*

\begin{proof}
Set $\gamma_j(t_0)=0$ for every job $j$ and time $t_0$. Subtracting
$t_0^q$ from \eqref{eq:capped-q-dual-certificate} and dividing by
$p_j>0$ gives
exactly the constraint of (DLP). Thus $(\alpha,\beta,\gamma)$ is a feasible
dual solution with objective value
\[
    \sum_{j=1}^n\alpha_j-s\int_0^\infty\sum_{d=1}^D\beta_d(t)\,\mathrm{d}t.
\]
By weak duality, this value is at most the optimal value of (LP).
By \cref{lem:capped-q-primal-relaxation}, that optimal value is at most
$\opt_{s,q}^q$, which proves the claim.
\end{proof}

\lemCappedQHeavyDualFitting*

\begin{proof}
We only put a positive price on the bottleneck resource $d_i$ of each
interval $I_i$. This price reflects the congestion $c_i$ in the interval and
the weight $\omega(T_i)$ at its end, and it applies to all times before
$T_i$. Formally, we define the following variables for the weak-duality
bound in \cref{lem:capped-q-dual-bound}.
\begin{alignat*}{2}
    \alpha_j&:=\sum_{i=1}^k a_{j,d_i}\min\{1,u_jc_i\}
        \int_{I_i\cap[0,C_j)}\omega(t)\,\mathrm{d}t
        &\quad&\text{for every }j\in[n],\\
    \beta_d(t)&:=\sum_{i\in[k]:\,t<T_i}
        \ind[d=d_i]c_i\omega(T_i)
        &&\text{for every }d\in[D],\ t\geq0.
\end{alignat*}
Every $\alpha_j$ is nonnegative because its defining sum has only
nonnegative terms. For a fixed resource $d$, $\beta_d(t)$ is a finite sum
of nonnegative terms. As $t$ increases, terms can only be removed from
this sum, so $\beta_d$ is nonincreasing. It is bounded and zero for
$t\geq T$, hence integrable.

It remains to verify \eqref{eq:capped-q-dual-certificate}. Fix a job $j$
and a time $t_0\geq0$. If $t_0\geq C_j$, disjointness of the intervals and
$a_{j,d_i}\min\{1,u_jc_i\}\leq1$ give
$\alpha_j\leq\int_0^{C_j}\omega(t)\,\mathrm{d}t=C_j^q\leq t_0^q$,
so the condition holds. Hence assume $t_0<C_j$. We have
\begin{align}
\alpha_j&=\sum_{i=1}^k a_{j,d_i}\min\{1,u_jc_i\}
    \int_{I_i\cap[0,C_j)}\omega(t)\,\mathrm{d}t \notag \\
    &= \sum_{i=1}^k a_{j,d_i}\min\{1,u_jc_i\}
        \int_{I_i\cap[0,t_0)}\omega(t)\,\mathrm{d}t
        +
        \sum_{i=1}^k a_{j,d_i}\min\{1,u_jc_i\}
            \int_{I_i\cap[t_0,C_j)}\omega(t)\,\mathrm{d}t \ .
            \label{eq:lq-heavy-1}
\end{align}
We first focus on the first sum of \eqref{eq:lq-heavy-1}.
Since the intervals $I_i$ are disjoint, $a_{j,d_i} \le 1$, and $\min\{1,u_jc_i\} \le 1$, we have 
\[
\sum_{i=1}^k a_{j,d_i}\min\{1,u_jc_i\}
    \int_{I_i\cap[0,t_0)}\omega(t)\,\mathrm{d}t
    \leq \int_0^{t_0}\omega(t)\,\mathrm{d}t=t_0^q
\]
by \eqref{eq:capped-q-weight}.
For the second sum in \eqref{eq:lq-heavy-1}, we have
\begin{align}
\sum_{i=1}^k a_{j,d_i}\min\{1,u_jc_i\}
    \int_{I_i\cap[t_0,C_j)}\omega(t)\,\mathrm{d}t
&=
\sum_{i\in[k]:\,t_0<T_i} a_{j,d_i}
    \int_{I_i\cap[t_0,C_j)}\min\{1,u_jc_i\}\omega(t)\,\mathrm{d}t
    \label{eq:lq-heavy-2} \\
&\leq
\sum_{i\in[k]:\,t_0<T_i} a_{j,d_i}\omega(T_i)
    \int_{I_i\cap[t_0,C_j)}\min\{1,u_jc_i\}\,\mathrm{d}t
    \label{eq:lq-heavy-3} \\
&\leq
\sum_{i\in[k]:\,t_0<T_i} a_{j,d_i}c_i\omega(T_i)
    \int_{I_i\cap[t_0,C_j)}x_j(t)\,\mathrm{d}t
    \label{eq:lq-heavy-4} \\
&\leq
\sum_{i\in[k]:\,t_0<T_i} a_{j,d_i}c_i\omega(T_i)
    \int_{t_0}^{C_j}x_j(t)\,\mathrm{d}t
    \label{eq:lq-heavy-5} \\
&\leq
p_j \sum_{i\in[k]:\,t_0<T_i} a_{j,d_i}c_i\omega(T_i)
    \label{eq:lq-heavy-6} \ .
\end{align}
The individual steps are justified as follows.
\begin{itemize}[nosep]
	\item \Cref{eq:lq-heavy-2} holds because $I_i\cap[t_0,C_j)$ is empty if
	        $t_0\geq T_i$.
	\item \Cref{eq:lq-heavy-3} uses that $\omega(\cdot)$ is nondecreasing.
	\item \Cref{eq:lq-heavy-4} uses that on any nonempty interval
	$I_i\cap[t_0,C_j)$, job $j$ is unfinished and $0<c(t)\leq c_i$. Hence
	$
    x_j(t)c_i
    =c_i\min\{u_j,1/c(t)\}
    \geq\min\{u_jc_i,1\}.
	$
	\item \Cref{eq:lq-heavy-5} uses that the integrand is nonnegative.
	\item \Cref{eq:lq-heavy-6} uses that job $j$ has at most $p_j$ units of
	processing left at time $t_0$.
\end{itemize}

Thus, by applying both upper bounds to \eqref{eq:lq-heavy-1}, we have in total
\[
\alpha_j \leq t_0^q
    +p_j\sum_{i\in[k]:\,t_0<T_i}a_{j,d_i}c_i\omega(T_i)
=t_0^q+p_j\sum_{d=1}^D a_{jd}\beta_d(t_0) \ ,
\]
which verifies the condition of \Cref{lem:capped-q-dual-bound}.
Therefore, the lemma implies that
\begin{equation}
\opt_{s,q}^q
\geq\sum_{j=1}^n\alpha_j-s\int_0^\infty\sum_{d=1}^D\beta_d(t)\,\mathrm{d}t \ . \label{eq:lq-heavy-8}
\end{equation}
It remains to compare the right-hand side of this inequality with $Z_q$.
To this end, define
\[
    S_q:=\sum_{i=1}^k c_i\int_{I_i}\omega(t)\,\mathrm{d}t
        =\sum_{i=1}^k c_i(T_i^q-T_{i-1}^q).
\]
Summing $\alpha_j$ over all jobs and applying
\eqref{eq:capped-q-selected-load} at each time in each interval gives
\begin{align*}
    \sum_{j=1}^n\alpha_j
    =\sum_{i=1}^k\int_{I_i}\omega(t)
        \sum_{j\in U(t)}a_{j,d_i}\min\{1,u_jc_i\}\,\mathrm{d}t
    \geq\frac12\sum_{i=1}^k c_i\int_{I_i}\omega(t)\,\mathrm{d}t
     =\frac12S_q.
\end{align*}
The definition of $\beta_d(t)$ gives
\begin{align*}
    \int_0^\infty\sum_{d=1}^D\beta_d(t)\,\mathrm{d}t
    &=\int_0^\infty\sum_{d=1}^D
        \sum_{i\in[k]:\,t<T_i}\ind[d=d_i]c_i\omega(T_i)\,\mathrm{d}t\\
    &=\sum_{i=1}^k c_i\omega(T_i)
        \int_0^{T_i}\sum_{d=1}^D\ind[d=d_i]\,\mathrm{d}t
    =\sum_{i=1}^k c_i\omega(T_i)T_i
    =q\sum_{i=1}^k c_iT_i^q \ ,
\end{align*}
where we use that 
$\sum_{d=1}^D\ind[d=d_i]=1$ and $\omega(T_i)=qT_i^{q-1}$.
Since $T_0=0$ and $c_{i+1}\leq c_i/2$, we have
\[
    \sum_{i=1}^k c_iT_{i-1}^q
    =\sum_{i=1}^{k-1}c_{i+1}T_i^q
    \leq\frac12\sum_{i=1}^k c_iT_i^q.
\]
Using the definition of $S_q$ and the preceding inequality, we obtain
\begin{align*}
    S_q
    =\sum_{i=1}^k c_i\bigl(T_i^q-T_{i-1}^q\bigr)
    =\sum_{i=1}^k c_iT_i^q
      -\sum_{i=1}^k c_iT_{i-1}^q
    \geq\sum_{i=1}^k c_iT_i^q
      -\frac12\sum_{i=1}^k c_iT_i^q
     =\frac12\sum_{i=1}^k c_iT_i^q.
\end{align*}
Thus $\int_0^\infty\sum_{d=1}^D\beta_d(t)\,\mathrm{d}t
=q\sum_{i=1}^k c_iT_i^q
\leq2qS_q$.
Together with \eqref{eq:lq-heavy-8} and setting $s=1/(8q)$ we have
\[
    \opt_{s,q}^q
    \geq\sum_{j=1}^n\alpha_j-s\int_0^\infty\sum_{d=1}^D\beta_d(t)\,\mathrm{d}t
    \geq\frac12S_q-\frac14S_q=\frac14S_q.
\]
At a heavy time, the definition gives $m(t)\leq Bc(t)$. Moreover,
$c(t)\leq c_i$ for every $t\in I_i$. Since $m(t)=0$ for $t\geq T$, only
the intervals $I_1,\ldots,I_k$ contribute to $Z_q^{\mathrm H}$. Therefore,
\begin{align*}
    Z_q^{\mathrm H}
    &=\sum_{i=1}^k
        \int_{I_i\cap\{t:\,t\text{ is heavy}\}}
            \omega(t)m(t)\,\mathrm{d}t
    \leq B\sum_{i=1}^k
        \int_{I_i\cap\{t:\,t\text{ is heavy}\}}
            \omega(t)c(t)\,\mathrm{d}t\\
    &\leq B\sum_{i=1}^k c_i
        \int_{I_i\cap\{t:\,t\text{ is heavy}\}}
            \omega(t)\,\mathrm{d}t
    \leq B\sum_{i=1}^k c_i\int_{I_i}\omega(t)\,\mathrm{d}t
     =BS_q.
\end{align*}
Since $Z_q^{\mathrm H}\geq Z_q^{\mathrm L}$, we conclude that
\[
    Z_q\leq2Z_q^{\mathrm H}\leq2BS_q\leq8B \cdot \opt_{s,q}^q.
\]
Finally, \cref{lem:capped-q-effective-decomposition} gives
\[
    \alg_q\leq\opt_{s,q}+Z_q^{1/q}
    \leq\bigl(1+(8B)^{1/q}\bigr)\opt_{s,q}.
\]
Substituting $s=1/(8q)$ proves the claimed bound.
\end{proof}

\lemCappedQLightDualFitting*

\begin{proof}
Define the following variables for the weak-duality bound in \Cref{lem:capped-q-dual-bound}.
\begin{alignat*}{2}
    \alpha_j&:=\int_{\{t:\,t\text{ is light}\}}
        \omega(t)z_j(t)\,\mathrm{d}t
        &\quad&\text{for every }j\in[n],\\
    \beta_d(t)&:=\frac{M(t)}B
        &&\text{for every }d\in[D],\ t\geq0.
\end{alignat*}
Every $\alpha_j$ is nonnegative because its defining integral has a
nonnegative integrand. Since $M(t)$ is nonnegative and nonincreasing, the
same is true of every $\beta_d(t)=M(t)/B$. The function $M$ is bounded and
zero after $m$ becomes zero.

It remains to verify \eqref{eq:capped-q-dual-certificate}. Fix a job $j$
and a time $t_0\geq0$. Since $z_j(t)\leq1$, the contribution to
$\alpha_j$ from times before $t_0$ is at most
$\int_0^{t_0}\omega(t)\,\mathrm{d}t=t_0^q$ by \eqref{eq:capped-q-weight}.
If $t_0\geq C_j$, there is no later contribution. Otherwise, at every
light time $t\in[t_0,C_j)$, job $j$ is unfinished and
\[
    \omega(t)z_j(t)
    =x_j(t)c(t)\omega(t)
    \leq x_j(t)\frac{m(t)\omega(t)}B
    \leq x_j(t)\frac{M(t_0)}B.
\]
The equality uses
\eqref{eq:capped-q-effective-rate}. The first inequality uses that $t$ is
light, so $c(t)<m(t)/B$. The second inequality follows from the definition
of $M(t_0)$, since $t\geq t_0$ implies
$\omega(t)m(t)\leq M(t_0)$.
Consequently,
\[
    \alpha_j
    \leq t_0^q+\frac{M(t_0)}B\int_{t_0}^{C_j}x_j(t)\,\mathrm{d}t
    \leq t_0^q+p_j\frac{M(t_0)}B
    \leq t_0^q+p_j\sum_{d=1}^D a_{jd}\beta_d(t_0).
\]
The second inequality uses that the remaining processing is at most $p_j$.
The last inequality uses $\sum_{d=1}^D a_{jd}\geq1$ because $\|a_j\|_\infty =1$ and demands are nonnegative. This verifies the condition of \cref{lem:capped-q-dual-bound}. Thus, the lemma implies that
\begin{equation}
\opt_{s,q}^q
\geq\sum_{j=1}^n\alpha_j-s\int_0^\infty\sum_{d=1}^D\beta_d(t)\,\mathrm{d}t \ . \label{eq:lq-light-1}
\end{equation}

It remains to compare the right-hand side of this inequality with $Z_q$.

To this end, we first prove the auxiliary bound
$\int_0^\infty M(t)\,\mathrm{d}t\leq qZ_q$.
Let $0<t_1<\cdots<t_h$ be the times at which $m$ decreases, including its
last decrease to zero, and let $\delta_i:=m(t_i^-)-m(t_i)\geq0$ be the size
of the decrease at time $t_i$. Since $m$ is nonincreasing and eventually
zero, its value at any time equals the sum of all its future decreases,
that is,
\[
    m(t_0)=\sum_{i\in[h]:\,t_0<t_i}\delta_i.
\]
Fix $t\geq0$ and consider any $t_0\geq t$. Since $\omega$ is
nondecreasing,
\begin{align*}
    \omega(t_0)m(t_0)
    =\sum_{i\in[h]:\,t_0<t_i}\delta_i\omega(t_0)
    \leq\sum_{i\in[h]:\,t_0<t_i}\delta_i\omega(t_i)
    \leq\sum_{i\in[h]:\,t<t_i}\delta_i\omega(t_i).
\end{align*}
The last expression is independent of $t_0$. Taking the supremum over
$t_0\geq t$ therefore gives
\[
    M(t)\leq\sum_{i\in[h]:\,t<t_i}\delta_i\omega(t_i).
\]
The term indexed by $i$ on the right is present precisely when $0\leq
t<t_i$. Integrating this bound gives
\begin{align*}
    \int_0^\infty M(t)\,\mathrm{d}t
    \leq\sum_{i=1}^h\delta_i t_i\omega(t_i)
    =q\sum_{i=1}^h\delta_i t_i^q
    &=q\sum_{i=1}^h\delta_i\int_0^{t_i}\omega(t)\,\mathrm{d}t\\
    &=q\int_0^\infty\omega(t)
        \sum_{i\in[h]:\,t<t_i}\delta_i\,\mathrm{d}t\\
    &=q\int_0^\infty\omega(t)m(t)\,\mathrm{d}t
     =qZ_q.
\end{align*}

We now bound the two terms on the right-hand side of
\eqref{eq:lq-light-1}. First, summing the definition of $\alpha_j$ over all
jobs gives
\begin{align*}
    \sum_{j=1}^n\alpha_j
    &=\sum_{j=1}^n
        \int_{\{t:\,t\text{ is light}\}}\omega(t)z_j(t)\,\mathrm{d}t\\
    &=\int_{\{t:\,t\text{ is light}\}}\omega(t)
        \sum_{j=1}^n z_j(t)\,\mathrm{d}t
    =\int_{\{t:\,t\text{ is light}\}}\omega(t)m(t)\,\mathrm{d}t
     =Z_q^{\mathrm L}
    \geq\frac12Z_q.
\end{align*}
The last inequality follows from the case assumption
$Z_q^{\mathrm L}\geq Z_q^{\mathrm H}$ and the identity
$Z_q=Z_q^{\mathrm H}+Z_q^{\mathrm L}$.

Second, since $\beta_d(t)=M(t)/B$ for every resource $d\in[D]$, we have
\begin{align*}
    \int_0^\infty\sum_{d=1}^D\beta_d(t)\,\mathrm{d}t
    =\int_0^\infty\sum_{d=1}^D\frac{M(t)}B\,\mathrm{d}t
    =\frac DB\int_0^\infty M(t)\,\mathrm{d}t
    \leq\frac{qD}{B}Z_q.
\end{align*}
The last inequality uses the bound
$\int_0^\infty M(t)\,\mathrm{d}t\leq qZ_q$ proved above.
We now set $s=B/(4qD) \in (0,1]$.
Together with \eqref{eq:lq-light-1}, the above gives
\[
    \opt_{s,q}^q
    \geq\sum_{j=1}^n\alpha_j-s\int_0^\infty\sum_{d=1}^D\beta_d(t)\,\mathrm{d}t
    \geq\frac12Z_q-\frac14Z_q=\frac14Z_q.
\]
Thus $Z_q\leq4\opt_{s,q}^q$. Finally,
\cref{lem:capped-q-effective-decomposition} gives
\[
    \alg_q\leq\opt_{s,q}+Z_q^{1/q}
    \leq\bigl(1+4^{1/q}\bigr)\opt_{s,q}.
\]
Substituting $s=B/(4qD)$ proves the claimed bound.
\end{proof}

\subsection{Lower Bounds}
\label{sec:lq-lower-bounds}

We now show that the minimum of the two upper bounds is tight up to constant
factors.
All jobs in the constructions below are released at time zero and have unit
rate caps. The first construction uses binary demands and applies to one-shot
DRF. It arranges jobs in blocks of disjoint resources, with geometrically
decreasing populations and geometrically increasing processing requirements.
Within each block, the instance resembles the one from
\cref{prop:drf-lower-bound}. DRF shares the resources equally between jobs
that use a single resource and jobs that use the entire block, whereas it
would be better to complete the single-resource jobs first. Across blocks,
one-shot DRF gives all jobs the same rate, so it is always slowed down by the
most congested unfinished block. In contrast, a comparison schedule processes
all blocks in parallel. For full DRF, we add small demands on earlier blocks
so that the same bottleneck freezes every unfinished job.

\begin{theorem}
\label{thm:lq-lower-bounds}
For every integer $D\geq1$ and real $q\geq1$, one-shot DRF and full DRF have a
competitive ratio in
\[
    \Omega\bigl(\min\{qD^{1/(q+1)},D\}\bigr)
\]
for minimizing the $\ell_q$ norm of completion times in multidimensional
scheduling with $D$ resources. The lower bound holds with
normalized demands, zero-release times, and unit rate caps. For one-shot DRF, it holds
even for binary demands.
\end{theorem}

\begin{proof}
\emph{Instance for one-shot DRF.}
Fix $D \ge 1$ and $q \ge 1$. We first construct a binary-demand instance for one-shot
DRF. Set $L:=\min\{D,\lfloor q\rfloor\}$
and $W:=\left\lfloor\frac DL\right\rfloor$.
Divide $L \cdot W$ resources into disjoint blocks $R_1,\ldots,R_L$, each containing
$W$ resources. The remaining resources are not used. Next, define
$\zeta:=W^{1/(q+1)}$, $K:=\lceil\zeta\rceil-1$, and $n_h:=2^{L-h}$ for $h\in[L]$.
For every block $R_h$, we create two groups of jobs.
\begin{itemize}[nosep]
    \item For each resource $d\in R_h$, create $n_h$ jobs that demand only
    resource $d$.
    \item Create $n_hK$ jobs that demand every resource in $R_h$ and no
    resource outside $R_h$.
\end{itemize}
Every positive demand is one, and every job associated with block $R_h$ has
processing requirement $p_h:=\frac{1}{n_h(K+1)}$.
Thus the number of jobs per resource halves from one block to the next, while
the processing requirements double. If $K=0$, the second group is empty.
On every resource, there are $K$ times as many block-wide jobs as
single-resource jobs. For $q=1$, there is only one block, and the instance
is similar to the one from \cref{prop:drf-lower-bound}.

\emph{Schedule of one-shot DRF.}
Suppose blocks $R_h,\ldots,R_L$ are unfinished. Each resource in $R_h$ is
demanded by $n_h(K+1)$ unfinished jobs, while every resource in a later block
is demanded by fewer jobs. Hence one-shot DRF assigns every unfinished job
rate $1/(n_h(K+1))$. Since all unfinished jobs receive the same rate and
$p_1<\cdots<p_L$, the blocks complete in increasing order of their indices.
This confirms that the unfinished jobs always form blocks
$R_h,\ldots,R_L$ for some $h$.

The first block completes at time one. When block $R_{h-1}$ completes, every
job in block $R_h$ has already received $p_{h-1}$ units of processing. For
$h\geq2$, the additional time needed to complete block $R_h$ is therefore
\[
    n_h(K+1)(p_h-p_{h-1})=\frac12.
\]
Consequently, all jobs in block $R_h$ complete at time $T_h=\frac{h+1}{2}$.
In particular, the last block contains $W+K$ jobs, all of which complete at
time $(L+1)/2$.

\emph{Comparison schedule.}
We compare this outcome with a schedule that processes all blocks in parallel.
In every block $R_h$, first run the single-resource jobs at rate $1/n_h$ for
$1/(K+1)$ time units. Each resource has load one, and all these jobs complete
at time $1/(K+1)$. If $K>0$, then run every remaining job in the block at rate
$1/(n_hK)$ for $K/(K+1)$ time units. Each resource again has load one, and
these jobs complete at time one. The blocks use disjoint resources, so these
steps can be performed simultaneously in all blocks. All rates are at most
one, and the schedule is feasible. It follows that
\begin{align*}
    \opt_{1,q}^q
    \leq
    \left(\frac{W}{(K+1)^q}+K\right)\sum_{h=1}^L n_h
    \leq 2\zeta(2^L-1),
\end{align*}
where the second inequality uses $K+1\geq\zeta$, $K\leq\zeta$,
$W=\zeta^{q+1}$, and $\sum_{h=1}^Ln_h=2^L-1$.

\emph{Bounding the ratio.}
Counting only the $W$ single-resource jobs in the last block in the
algorithm's objective gives
\begin{align*}
    \frac{\alg_q}{\opt_{1,q}}
    \geq\frac{L+1}{2}
        \left(\frac{W}{2\zeta(2^L-1)}\right)^{1/q}
    \geq\frac{L+1}{2\cdot2^{(L+1)/q}}W^{1/(q+1)}
    \geq\frac{L}{8}W^{1/(q+1)}.
\end{align*}
The last inequality follows from $L\leq q$ and $q\geq1$, which imply
$2^{(L+1)/q}\leq4$. If $q\leq D$, then
$L\geq q/2$ and $W\geq D/(2q)$. Since
$(2q)^{1/(q+1)}\leq2$, the ratio is at least
$qD^{1/(q+1)}/32$. If $q>D$, then $L=D$ and $W=1$, and the ratio is at least
$D/8$. Thus the claimed asymptotic lower bound holds for one-shot DRF.

\emph{Extension to full DRF.}
The instance above does not directly apply to full DRF. When $L>1$, full
DRF continues to increase the rates of jobs in later blocks after the
resources of the first unfinished block become tight, because these jobs use
disjoint resources. We now add small cross-block demands to prevent this
redistribution. We keep the processing requirements and the original
demands. Additionally, we give every job associated with block $R_\ell$ a
demand of $\varepsilon:=\frac{1}{4LW}$ on each resource in every earlier
block $R_h$, where $h<\ell$. Every job still has demand one on some resource
in its own block, so all demand vectors remain normalized.

Suppose blocks $R_i,\ldots,R_L$ are unfinished. Let $w_h^{(i)}$ denote their
total demand on a resource in block $R_h$. By symmetry, this quantity is the
same for every resource in the block. Let $N:=W+K$. Since block $R_\ell$
contains $n_\ell N$ jobs, we have
\[
    w_h^{(i)}
    =\ind[h\geq i]n_h(K+1)
     +\varepsilon N
        \sum_{\ell=\max\{h+1,i\}}^L n_\ell.
\]
In particular, $w_i^{(i)}\geq n_i(K+1)$. Moreover,
$N\leq W(K+1)$ implies $\varepsilon N\leq(K+1)/4$. Using
$n_{h+1}=n_h/2$ and $\sum_{\ell=h+1}^L n_\ell<n_h$, we obtain
\begin{alignat*}{3}
    w_h^{(i)}
    &\leq2\varepsilon N n_i
     \leq\frac12n_i(K+1)
        &\quad&\text{for }h<i,\\
    w_h^{(i)}
    &\leq n_h(K+1+\varepsilon N)
     \leq\frac58n_i(K+1)
        &&\text{for }h>i.
\end{alignat*}
Thus the resources in $R_i$ carry the largest total demand and are the first
to reach capacity when the rates of all unfinished jobs increase together. Every unfinished job demands at least
one of these resources, so full DRF freezes every job during the initial round of
progressive filling, at the common rate $1/w_i^{(i)}$.

The unfinished jobs again receive equal rates and the blocks complete in
increasing order. Set $p_0:=0$. The time between the completions of blocks
$R_{i-1}$ and $R_i$ is $w_i^{(i)}(p_i-p_{i-1})$, where the first interval
starts at time zero. Therefore, the completion time of block $R_h$ satisfies
\[
    T_h
    =\sum_{i=1}^h w_i^{(i)}(p_i-p_{i-1})
    \geq\sum_{i=1}^h n_i(K+1)(p_i-p_{i-1})
    =\frac{h+1}{2}.
\]
Hence the algorithm's completion times are no smaller than those in the
one-shot construction.

It remains to adapt the comparison schedule to the new cross-block demands.
During its first step, the jobs from block $R_h$ create load one on each
resource in that block, while the single-resource jobs from each later block
add load $\varepsilon W$. Thus the load on a resource in $R_h$ is at most
\[
    1+\varepsilon W(L-h)\leq\frac54.
\]
If $K>0$, then during the second step each later block adds load
$\varepsilon$, so the load is at most
\[
    1+\varepsilon(L-h)\leq\frac54.
\]
Scale every rate in the comparison schedule by $4/5$ and every duration by
$5/4$. The resulting schedule is feasible and its completion-time norm is
$5/4$ times the norm of the original comparison schedule. Since the
algorithm's completion times have not decreased, its ratio is at least
$4/5$ times the one-shot lower bound, and hence at least
\[
    \frac1{40}\min\{qD^{1/(q+1)},D\}.
\]
This proves the theorem.
\end{proof}

\section{A Black-Box Reduction for Release Times}
\label{sec:release-time-reduction}

The zero-release analyses in
\cref{sec:completion-time,sec:low-dimensional-bounds,sec:lq-general-case}
use that the unfinished set only shrinks when all jobs are released at time
zero. With arbitrary release times, new arrivals may decrease the rates of
existing jobs. We now show that monotonicity with respect to the active set
still suffices to extend these guarantees, losing only a factor of two.
The reduction is black-box in the sense that it applies to every allocation
rule with a few natural properties, not only to one-shot DRF.
We then restate the resulting bounds for total completion time and
$\ell_q$ norms with arbitrary release times.

A \emph{set-based} allocation rule assigns to every set $A$ of jobs a
feasible rate vector $x(A)$ with $x_j(A)=0$ for $j\notin A$. The rule may depend on
the demand vectors and rate caps of the jobs in $A$, but not on their
processing requirements or release times. Over time, it runs the jobs at
rates $x(A(t))$. In other words, the allocation depends on the execution only
through the active set. We call the rule \emph{positive} if
$x_j(A)>0$ for every $j\in A$, and \emph{monotone} if $A\subseteq B$
implies $x_j(A)\geq x_j(B)$ for every $j\in A$. We state the reduction for
the $\ell_q$ norm of completion times with $q\geq1$, which includes total
completion time as the special case $q=1$.

\begin{theorem}
\label{thm:release-time-reduction}
Let $q\geq1$, and consider a positive, monotone, set-based allocation rule
for multidimensional scheduling. If the rule is $\mu$-competitive for
minimizing the $\ell_q$ norm of completion times with zero-release times,
then it is $2\mu$-competitive for this objective with arbitrary release
times $r_j\geq0$.
\end{theorem}

The idea is to construct a virtual instance in which all jobs are released
at time zero and on which the rule produces exactly the same completion
times. Each virtual job may receive processing before its original release
time, while its rate after release can be lower than in the original
execution. Its virtual processing requirement is therefore at most its
original requirement plus the virtual work received before release.
We bound the optimum of the virtual instance by combining this pre-release work
with an optimal schedule for the original instance. Running both in
parallel at half speed preserves feasibility and at most doubles every
completion time.

\begin{proof}
Fix an instance with processing requirements $p$ and release-time vector $r$,
and run the allocation rule on this instance. Let $C_j$ be the resulting
completion time of job $j$, and write $C:=(C_j)_{j=1}^n$. Denote the
$\ell_q$ norms of the completion-time vectors of the rule and of an optimal
schedule by $\alg_r(p)$ and $\opt_r(p)$, respectively. Since demand vectors
and rate caps stay fixed throughout the proof, we omit them from this
notation. A subscript $0$ denotes the corresponding values when all release
times are zero. At time $t$, let $A(t):=\{j:r_j\leq t<C_j\}$ be
the active set and let $U(t):=\{j:t<C_j\}$ contain all jobs that have not yet
completed, including those not yet released. Since $A(t)\subseteq U(t)$,
monotonicity gives $x_j(A(t))\geq x_j(U(t))$ for every $j\in A(t)$.

We associate with the release-time instance a virtual instance in which all
jobs are available at time zero. The virtual processing requirement of job
$j$ is
\[
    \hat p_j:=\int_0^{C_j}x_j(U(t))\,\mathrm{d}t.
\]
This instance is used only for the analysis and need not be constructed
online. It has the same demand vectors and rate caps, so whenever the
virtual active set is $U(t)$, the rule assigns the rate vector $x(U(t))$. The definition of $\hat p_j$ makes
job $j$ complete at time $C_j$. Positivity rules out an earlier
completion, since for every $0\leq t<C_j$, we have
\[
    \hat p_j-\int_0^t x_j(U(s))\,\mathrm{d}s
    =\int_t^{C_j}x_j(U(s))\,\mathrm{d}s>0.
\]
An induction over the distinct completion times therefore shows that the
virtual execution has active set $U(t)$ and the same completion-time vector
$C$ as the original execution. Hence $\alg_r(p)=\alg_0(\hat p)$.

It remains to compare the two optimal values. Define the virtual work
accumulated before the release of job $j$ by
$b_j:=\int_0^{r_j}x_j(U(t))\,\mathrm{d}t$. The rate comparison above implies
\begin{equation}
\label{eq:virtual-processing-decomposition}
    \hat p_j
    =b_j+\int_{r_j}^{C_j}x_j(U(t))\,\mathrm{d}t
    \leq b_j+\int_{r_j}^{C_j}x_j(A(t))\,\mathrm{d}t
    =b_j+p_j.
\end{equation}

Let $x^*$ be an optimal schedule for the original instance, with completion
times $C_j^*$, and write $C^*:=(C_j^*)_{j=1}^n$. To process the virtual
instance, we combine the pre-release work with the processing in $x^*$.
Consider the rates
\[
    y_j(s):=\frac12 x_j(U(s/2))\ind[s/2<r_j]
        +\frac12 x_j^*(s/2).
\]
In words, $y$ runs at half speed. It uses one half of the capacity to
perform the virtual work $b_j$ that each job receives before its release,
and the other half to follow $x^*$. Since all demands are nonnegative,
setting some coordinates of a feasible rate vector to zero keeps it
feasible. In particular, the vector obtained from $x(U(s/2))$ by setting the
coordinates of released jobs to zero is feasible. Moreover, the rate caps
and resource capacities are linear constraints, so the average $y(s)$ of two
feasible rate vectors is feasible as well. Finally,
$r_j\leq C_j^*$ and a change of variables give
\begin{align*}
    \int_0^{2C_j^*}y_j(s)\,\mathrm{d}s
    &=\int_0^{r_j}x_j(U(t))\,\mathrm{d}t
      +\int_0^{C_j^*}x_j^*(t)\,\mathrm{d}t \\
    &=b_j+p_j
    \geq\hat p_j,
\end{align*}
where the last inequality follows from
\eqref{eq:virtual-processing-decomposition}. Thus every virtual job $j$
receives its required processing by time $2C_j^*$. Any processing after a
virtual job completes may again be set to zero without violating
feasibility. Since the $\ell_q$ norm is monotone in every coordinate, we
obtain $\opt_0(\hat p)\leq\lVert2C^*\rVert_q=2\opt_r(p)$. Consequently,
\[
    \alg_r(p)
    =\alg_0(\hat p)
    \leq\mu\opt_0(\hat p)
    \leq2\mu\opt_r(p),
\]
which proves the theorem.
\end{proof}

One-shot DRF is set-based because its allocation only depends on the demand
vectors and rate caps of the active jobs. It is positive because $\lambda(A)\geq1/|A|>0$, and it is monotone by
\cref{obs:one-shot-monotonicity}. Hence, \cref{thm:release-time-reduction}
applies. We transfer the resulting guarantees to full DRF using
\cref{lem:full-drf-dominates-one-shot}, which holds for arbitrary release
times.

\begin{corollary}
\label{cor:drf-release-times}
For multidimensional scheduling with $D\geq1$ resources and arbitrary release times
$r_j\geq0$, one-shot DRF and full DRF are $(8\sqrt D+2)$-competitive for
minimizing total completion time. For $D=2$ and $D=3$, respectively, the
upper bounds improve to $4+2\sqrt3$ and
$4+2\sqrt7$. For both algorithms, the competitive ratio is
$\Theta(\sqrt D)$.
\end{corollary}

\begin{proof}
Apply \cref{thm:release-time-reduction} with $q=1$ to the one-shot DRF bounds in
\cref{thm:drf-completion-time,thm:drf-two-resources,thm:drf-three-resources},
and then use \cref{lem:full-drf-dominates-one-shot} for full DRF.
The matching lower bound from \cref{prop:drf-lower-bound} already holds
with zero-release times, so it also applies here.
\end{proof}

For the norm objective, write $\alg_q:=\lVert C\rVert_q$ for the norm of
the algorithm's completion-time vector $C$, and let $\opt_{1,q}$ be the
minimum norm over feasible unit-speed schedules. Both values refer to the
original instance with its given release times.

\begin{corollary}
\label{cor:drf-norms-release-times}
For every integer $D\geq1$ and real $q\geq1$, one-shot DRF and full DRF
satisfy
\[
    \alg_q\leq\min\bigl\{144qD^{1/(q+1)},4D+2\bigr\}\opt_{1,q}
\]
on every instance of multidimensional scheduling with $D$ 
resources and arbitrary release times. Their competitive ratios for minimizing the
$\ell_q$ norm of completion times are in
\[
    \Theta\bigl(\min\{qD^{1/(q+1)},D\}\bigr) \ .
\]
In particular, both
algorithms have competitive ratio $\Theta(D)$ when $q\geq D$.
\end{corollary}

\begin{proof}
The proofs of \cref{thm:lq-q-upper-bound,thm:lq-dimension-upper-bound}
give one-shot DRF competitive factors $72qD^{1/(q+1)}$ and $2D+1$,
respectively, when all jobs are released at time zero. Applying
\cref{thm:release-time-reduction} to each bound gives the displayed upper
bound for arbitrary release times. Since the
reduction acts on the norm itself, rather than on the sum of $q$th powers,
its factor of two is independent of $q$. By
\cref{lem:full-drf-dominates-one-shot}, full DRF completes every job no
later than one-shot DRF on the same instance, so it satisfies the same
norm bound.

The lower bound in \cref{thm:lq-lower-bounds} uses only instances with
zero-release times and unit rate caps, and hence also applies to the
larger class of instances considered here. Together with the upper
bound, it gives the claimed $\Theta$ bound.
\end{proof}

\end{document}